\documentclass[useAMS,referee,usenatbib]{biom}

\usepackage{rotating,tabularx,listings} 
\usepackage{url}
\usepackage{algorithm} 
\usepackage{algpseudocode} 
\usepackage{amsmath}
\usepackage{amssymb}
\usepackage[mathscr]{euscript}
\usepackage[colorlinks=true,linkcolor=red,citecolor=blue]{hyperref}
\usepackage{url}
\usepackage{natbib}
\usepackage{multirow}
\usepackage{booktabs}
\usepackage{hhline}
\usepackage{enumerate}
\usepackage{verbatim}
\usepackage{dsfont}
\usepackage[colorlinks=true,linkcolor=red,citecolor=blue]{hyperref}
\usepackage{threeparttable}
\usepackage{array}
\usepackage{float}
\usepackage{xcolor}
\usepackage{subcaption}

\usepackage{xcolor}
\usepackage{soul}
\usepackage{tikz}
\usetikzlibrary{calc}

\newtheorem{proposition}{Proposition}

\newcommand{\bzero}{\boldsymbol{0}}
\newcommand{\bone}{\boldsymbol{1}}

\newcommand{\bY}{\boldsymbol{Y}}
\newcommand{\bC}{\boldsymbol{C}}
\newcommand{\bX}{\boldsymbol{X}}
\newcommand{\bR}{\boldsymbol{R}}
\newcommand{\bO}{\boldsymbol{O}}
\newcommand{\bW}{\boldsymbol{W}}
\newcommand{\bw}{\boldsymbol{w}}
\newcommand{\bB}{\boldsymbol{B}}

\newcommand{\bE}{\boldsymbol{E}}

\newcommand{\bpsi}{\boldsymbol{\psi}}
\newcommand{\btheta}{\boldsymbol{\theta}}

\newcommand{\bfB}{\mathbf{B}}
\newcommand{\bfO}{\mathbf{O}}

\def\bSig\mathbf{\Sigma}

\title[Comprehensively handling missing data in CRTs]{Handling incomplete outcomes and covariates in cluster-randomized trials: doubly-robust estimation, efficiency considerations, and sensitivity analysis}

\author{Bingkai Wang$^{1,*}$\email{bingkai.w@gmail.com}, 
Fan Li$^{2,3}$, and Rui Wang$^{4,5}$\\
$^{1}$Department of Biostatistics, School of Public Health, University of Michigan, Ann Arbor, MI, U.S.A.\\
$^{2}$Department of Biostatistics, Yale School of Public Health, New Haven, CT, U.S.A. \\
$^{3}$Center for Methods in Implementation and Prevention Science, Yale School of Public Health,\\ 
New Haven, CT, U.S.A.\\
$^{4}$Department of Population Medicine, Harvard Pilgrim Health Care Institute and Harvard Medical School, \\
Boston, MA, U.S.A.\\
$^{5}$Department of Biostatistics, Harvard T.H. Chan School of Public Health, Boston, MA, U.S.A.
}

\begin{document}





\pagerange{\pageref{firstpage}--\pageref{lastpage}} 
\volume{64}
\pubyear{2008}
\artmonth{December}


\doi{}


\label{firstpage}


\begin{abstract}
In cluster-randomized trials (CRTs), missing data can occur in various ways, including missing values in outcomes and baseline covariates at the individual or cluster level, or completely missing information for non-participants. Among the various types of missing data in CRTs, missing outcomes have attracted the most attention. However, no existing methods simultaneously address all aforementioned types of missing data in CRTs. To fill in this gap, we propose a doubly-robust estimator for the average treatment effect on a variety of effect measure scales. The proposed estimator simultaneously handles missing outcomes under missingness at random, missing covariates without constraining the missingness mechanism, and missing cluster-population sizes via a uniform sampling mechanism. Furthermore, we detail key considerations to improve precision by specifying the optimal weights, leveraging machine learning, and modeling the treatment assignment mechanism. Finally, to evaluate the impact of violating missing data assumptions, we contribute a new sensitivity analysis framework tailored to CRTs. We assess the performance of the proposed methods through simulations and illustrate their use in a real data application. 
\end{abstract}



\maketitle

\section{Introduction}
\begin{figure}[ht]
    \centering
    \includegraphics[width=0.8\textwidth]{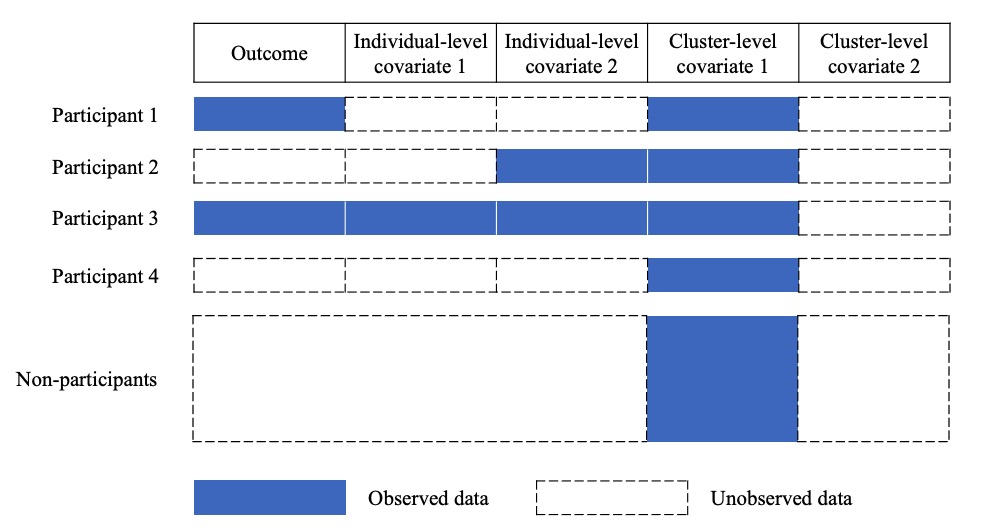}
    \caption{A schematic illustration of missing data in a cluster of a hypothetical CRT with 4 observed (enrolled) participants. }
    \label{fig:1}
\end{figure}

In a cluster-randomized trial (CRT), groups of individuals, such as classrooms, hospitals, or villages, are randomized, and the within-cluster observations are often correlated \citep{turner2017review}. 
Missing data are ubiquitous in CRTs. In a systematic review of CRTs published in 2011 \citep{diaz2014missing}, 95 out of 132 (72\%) CRTs involved missing data, and the missing data proportion ranged from 1\% to 47\%. However, 65 out of the 95 (66\%) CRTs adopted complete case analyses, ignoring missing data. 
Indeed, missing data can arise through multiple mechanisms and at various hierarchical levels in CRTs. At the individual level, participants may have missing values in both outcome and covariates. At the cluster level, beyond missing values in cluster-level covariates, some individuals within a cluster may not be enrolled, leading to missing individual-level information for non-participants; 
see Figure~\ref{fig:1} for a schematic illustration. 

Most prior literature on missing data in CRTs has focused on missing outcomes under the missing-at-random assumption. Multilevel multiple imputation \citep{taljaard2008imputation, diaz2014handling} is a model-based approach for filling in missing data, and its validity relies on a correctly specified imputation model \citep{turner2020properties}. \cite{prague2016accounting} combined augmented generalized estimating equations (Aug-GEE, \citealp{stephens2012augmented}) and inverse probability weighting (IPW) to analyze CRTs with missing outcomes, under the assumption of non-informative cluster sizes. A key feature of this estimator is its double-robustness, i.e., consistent estimation if either the outcome model or the missingness propensity score is correctly specified. However, in the presence of informative cluster size, standard GEE estimators for the average treatment effect may be biased \citep{wang2022two}.  
\cite{balzer2023two} and \citet{nugent2024blurring} considered the targeted maximum likelihood estimation (TMLE) to address missing outcomes; we discuss their approaches in detail in Section~\ref{sec: discussion}.

Despite prior efforts, methods remain lacking to jointly address missing data in outcomes, covariates, and non-enrolled participants in CRTs, as well as to conduct sensitivity analysis under violations of missing data assumptions in this context. To fill this gap, we build on the doubly robust inference framework \citep{bang2005doubly} and develop methods that (i) incorporate scaled weights to account for clustering and achieve doubly robustness under arbitrary correlation structures while accommodating informative cluster sizes, (ii) extend missing-indicator approaches for handling covariate missingness \citep{zhao2022adjust, chang2023covariate,zhao2023covariate} from individually randomized trials to CRTs, and (iii) introduce a framework for addressing non-participation through new identifiability assumptions and accompanying asymptotic theory.

Furthermore, we contribute three key insights into the practical implementation of our proposed estimators to maximize statistical power, through the derivation of optimal weights, the incorporation of debiased machine learning \citep{chernozhukov2018double} for estimating nuisance functions, and a formal evaluation of the potential efficiency gains from additionally including a propensity score model for treatment \citep{balzer2016targeted}. Finally, we introduce a new sensitivity analysis framework to assess potential biases from both selection bias and outcome missing-not-at-random mechanisms, based on computing a tipping point that characterizes the minimal divergence between the observed and missing data distributions required to overturn the primary conclusions.

\section{Notation and assumptions}\label{sec: notation}
We consider a CRT with $m$ clusters. Each cluster $i$ ($i=1,\dots, m$) contains $N_i$ individuals, but only $M_i$ individuals $(M_i\le N_i)$ are enrolled. Here, $N_i$ is called the \emph{cluster-population size} (the total number of individuals of interest), and $M_i$ is called the \emph{sampled cluster size} (the number of enrolled individuals).
For each individual $j$ ($j=1,\dots, N_i)$ in cluster $i$, we denote $Y_{ij}$ as the outcome, $\bX_{ij} = (X_{ij1},\dots, X_{ijp})^\top$ as a $p$-dimensional vector of individual-level baseline covariates, and {$E_{ij}$ as the enrollment indicator} such that $M_i = \sum_{j=1}^{N_i} E_{ij}$. We further let $A_i$ represent a binary cluster-level treatment (with 1 for treatment and 0 for control) and $\bC_i = (C_{i1}, \dots, C_{iq})^\top$ be a $q$-dimensional vector of cluster-level baseline covariates.

In practice, outcomes, covariates, and cluster-population sizes are all subject to missingness. We define $R_{ij}^Y$ as a non-missingness indicator of $Y_{ij}$  with $R_{ij}^Y = 1$ if $Y_{ij}$ is observed and $R_{ij}^Y = 0$ otherwise. Similarly, we define $R_{ijk}^X, R_{il}^C, R_i^N$ as the non-missingness indicator for covariates $X_{ijk}, C_{il}, N_i$, respectively $(k=1,\dots, p;l=1,\dots,q)$. 
We assume that $A_i$ and $M_i$ are always observed by design, but the enrollment index set $\{j: E_{ij}=1\}$ is unobserved if $M_i<N_i$, because CRTs often do not have data for nonparticipants.
To simplify the notation, we define $X_{ijk}^* = (R_{ijk}^X, R_{ijk}^XX_{ijk})$ as the observed information for $X_{ijk}$; in the special case $R_{ijk}^X \equiv 1$, we set $X_{ijk}^*=X_{ijk}$. Similarly, $C_{il}^* = (R_{il}^C, R_{il}^CC_{il})$ and $N_i^* = (R_i^N, R_i^NN_i)$ denote the observed cluster-level covariates and observed cluster-population size. Finally, the observed data for each cluster become . 

To begin with, we consider a basic setting where $M_i = N_i$, meaning that all individuals in a cluster are enrolled. 
In practice, this setting corresponds to scenarios where clusters represent household, class, or working group so that $N_i$ is relatively small.
In Sections~\ref{sec: est-sampling} and \ref{sec: sensitivity}, we will return to the general setting where $M_i\le N_i$. 
{This setting will correspond to CRTs with large clusters such as villages, hospitals, or cities, where the interest lies in generalizing the treatment effect estimated from the $M_i$ samples to all $N_i$ individuals.}
Under the basic setting with $M_i=N_i$ (meaning $R_i^{N}=1$), the observed data become $\bfO_i = (\bR_i^Y, \bR_i^Y \circ \bY_i, A_i, \bfB_i)$, where  $\bY_i= (Y_{i1}, \dots, Y_{iN_i})^\top$,  $ \bR_i^Y= (R_{i1}^Y, \dots, R_{iN_i}^Y)^\top$, and $\bfB_i = (\bB_{i1},\dots, \bB_{iN_i})^\top$ with ``$\circ$'' denoting element-wise product between vectors.
Furthermore, we define the outcome-complete data as $\bW_i = (\bR_i^Y, \bY_i, A_i, \bfB_i)$ and make the following structural assumptions on $\{\bW_1,\dots, \bW_m\}$.

\begin{assumption}[Super-population]\label{assp1}
(1) $\{\bW_1,\dots, \bW_m\}$ are independent and identically distributed samples from an unknown distribution $\mathcal{P}^{\bW}$ with finite second moments. (2) Within each cluster $i$, the individual-level data $\{(R_{ij}^Y, Y_{ij}, \bB_{ij}^\top): j = 1,\dots, N_i\}$ are identically distributed given $(A_i, N_i)$.
\end{assumption}

\begin{assumption}[Cluster randomization]\label{assp2}
In the distribution $\mathcal{P}^{\bW}$, $P(A_i=1) = \pi$ for some known constant $\pi \in (0,1)$, and $A_i$ is independent of $\bfB_i$.
\end{assumption}

\begin{assumption}[Restricted missing at random, rMAR]\label{assp3}
Each $R_{ij}^Y$ is conditionally independent of $\bY_i$ given $(A_i, \bfB_i)$ and $P(R_{ij}^Y=1|A_i, \bfB_i)$ is uniformly bounded away from 0.
\end{assumption}

Assumption \ref{assp1}(1) describes a super-population framework for inference and assumes inter-cluster independence. 
Assumption \ref{assp1}(2) requires that, given the treatment condition and cluster-population size, each individual in a cluster is identically distributed, but they can be arbitrarily correlated. 
Assumption \ref{assp2} characterizes cluster randomization and assumes that missingness in covariates is independent of treatment assignments. Of note, we allow for the baseline covariates to be missing not at random (MNAR).  
{While this setup reflects simple randomization, our methods also apply to more general designs (e.g., stratified randomization), as discussed in Section~\ref{sec: discussion}.}
Assumption~\ref{assp3}, like the classical MAR assumption,  induces conditional independence between missing-outcome indicators and outcomes given treatment and covariates. 
In the CRT setting, rMAR is {different from classical MAR in that it does not condition on observed outcomes within the same cluster; instead, it requires that the missingness of any outcome is independent of all outcomes in the cluster conditional on $(A_i, {\bf B}_i)$}. 
In Section~\ref{sec: sensitivity}, we provide sensitivity analysis for this rMAR assumption. 

Given Assumptions \ref{assp1}-\ref{assp3}, our goal is to estimate the average treatment effect, defined as 
\begin{equation}\label{eq:estimand}
        \Delta = f(E[Y_{ij}|A_i=1], E[Y_{ij}|A_i=0]),
\end{equation}
where $f$ determines the scale of effect measure, e.g., $f(x,y) = x-y$ for the difference estimand and $f(x,y) = x(1-y)/\{(1-x)y\}$ for the odds ratio estimand.
In the context of weighted average treatment effects in CRTs (\citealp{kahan2022estimands, benitez2021comparative} among others), Equation~(\ref{eq:estimand}) corresponds to the cluster-average treatment effect as $E[Y_{ij}|A_i=a]=E[N_i^{-1}\sum_{j=1}^{N_i}Y_{ij}|A_i=a]$ under Assumption \ref{assp1} for $a=0,1$. 


\section{Doubly-robust estimation with incomplete outcome and covariate data}\label{sec: dr-estimation}

We estimate $\Delta$ with the following steps. First, we construct an outcome regression model $\eta(A_i, {\bfB_i};\btheta_Y)$ for $E[Y_{ij}|R_{ij}=1,A_i, \bfB_{i}]$ and a propensity score function $\kappa(A_i, {\bfB_i};\btheta_R)$ for $P(R_{ij}^Y=1|A_i, \bfB_{i})$, where $\btheta_Y$ and $\btheta_R$ are nuisance parameters. 
For example, we can set a linear outcome model $\eta(A_i, {\bfB_i};\btheta_Y) = \beta_0 +\beta_AA_i + \boldsymbol{\beta}_{\bB}^\top \bB_{ij} + \boldsymbol{\beta}_{\bX^o}^\top \overline{\bX_i^o}$ with $\overline{\bX_i^o}$ being the cluster-average of observed individual-level covariates and $\btheta_{Y} = ( \beta_0, \beta_A,\boldsymbol{\beta}_{\bB}^\top,\boldsymbol{\beta}_{\bX^o}^\top)^\top$. 
We do not restrict the estimator to any specific working model, but only require $\btheta_Y$ and $\btheta_R$ are estimated by solving pre-specified estimating equations $\bpsi_Y(\bfO_i;\btheta_Y)$ and $\bpsi_R(\bfO_i;\btheta_R)$ so that $\sum_{i=1}^m \bpsi_Y(\bfO_i;\widehat{\btheta}_Y) = \bzero$ and $\sum_{i=1}^m \bpsi_R(\bfO_i;\widehat{\btheta}_R) = \bzero$. This setting covers a wide variety of estimators, such as maximum likelihood, penalized regression, {and imputation methods for missing covariates. For example, if missing values in $X_{ijk}$ are imputed by a constant $c$, then the imputed covariate is $R_{ijk}^XX_{ijk} + (1-R_{ijk}^X)c$, which is a function of $X_{ijk}^*$. More generally, when the imputed covariate vector remains a function of $\bfB_{i}$, this procedure can be incorporated into the estimating equations for nuisance parameters. }

In the next step, we define
\begin{equation}\label{eq: est}
 D_{ij}(a, \btheta_Y, \btheta_R) = \frac{w_j(N_i)}{\sum_{j'=1}^{N_i} w_{j'}(N_i)} \left[ \frac{I\{A_i=a\}}{\pi^a(1-\pi)^{1-a}}     \frac{R_{ij}^Y \{Y_{ij} - \eta(a, {\bfB_i};\btheta_Y)\}}{ \kappa(a, {\bfB_i};\btheta_R)} + \eta(a, {\bfB_i};\btheta_Y)\right]
\end{equation}
for $a = 0,1$ and arbitrary positive weights $w_j(N_i)$.
Then, we estimate $E[Y_{ij}|A_i=a]$ by $\widehat{\mu}_a = m^{-1} \sum_{i=1}^m \sum_{j=1}^{N_i} D_{ij}(a, \widehat{\btheta}_Y, \widehat{\btheta}_R)$ and compute $\widehat{\Delta} = f(\widehat\mu_1, \widehat\mu_0)$. In the final step, we construct a sandwich variance estimator for $\widehat{\Delta}$, denoted as $\widehat{Var}(\widehat{\Delta})$; 
{the details of variance estimation are provided in Supplementary Material A.}

To derive the asymptotic results, we assume standard regularity conditions, which are given in Supplementary Material B and similar to those involved in \cite{vaart_1998} for M-estimation. To proceed, we denote $\btheta_R^*, \btheta_Y^*$ as the unique solution to $E[\bpsi_Y(\bfO_i;\btheta_R)]=\bzero$ and $E[\bpsi_R(\bfO_i;\btheta_Y)]=\bzero$, hence representing the probability limit of $\widehat{\btheta}_R, \widehat{\btheta}_Y$. The following theorem summarizes the consistency and asymptotic normality of the proposed estimator.

\begin{theorem} \label{thm1}
    Given Assumptions~\ref{assp1}-\ref{assp3} and regularity conditions in Supplementary Material B, if either $\kappa(A_i, {\bfB_i};\btheta_R^*) = P(R_{ij}^Y=1|A_i, \bfB_{i})$  or $\eta(A_i, {\bfB_i};\btheta_Y^*)=E[Y_{ij}|R_{ij}=1,A_i,\bfB_{i}]$, then $\{\widehat{Var}(\widehat{\Delta})\}^{-1/2} (\widehat{\Delta} - \Delta) \xrightarrow{d} N(0,1)$. 
\end{theorem}

Theorem~\ref{thm1} establishes the double-robustness property, 
which has two important implications. 
First, double-robustness holds under an arbitrary correlation structure among $\bR_i^Y, \bY_i$, or $\bfB_i$; thus different decisions in modeling the correlation structure only affect the efficiency of $\widehat{\Delta}$ but not its consistency. Second, covariate missingness distributions have no impact on double-robustness. That is, under Assumptions~\ref{assp1}-\ref{assp3}, the missingness indicator approach is effective in handling MNAR in covariates in CRTs, 
extending the results in \cite{zhao2023covariate} on missing covariates in individually-randomized trials. 

\subsection{Key considerations for improving precision}
\subsubsection{Choosing the optimal weights}\label{sec:opt_weights}
The weights $\bw(N_i) =\{w_{1}(N_i), \dots, w_{N_i}(N_i)\}^\top$ correspond to the working correlation matrix in the standard GEE for CRTs. 
A common practice is to use an exchangeable correlation structure (i.e., $\bw(N_i) =  \textup{Corr}(\rho)^{-1} \bone_{N_i}$ with $\textup{Corr}(\rho)$ being the exchangeable correlation matrix in GEE and $\bone_{N_i}$ being a $N_i$-dimensional vector of ones), while others suggest adopting constant weights, $\bw(N_i) = \bone_{N_i}$ \citep{kahan2022estimands}. {Our derivation shows that, whichever weight we choose, it is important to offset its impact on modifying the estimand by adding a factor $1/\bw(N_i)^\top \bone_{N_i}$ such that we are not targeting $E[\bw(N_i)^\top\bone_{N_i}]^{-1}E[\bw(N_i)^\top \bY_i|A_i=a]$--- a different estimand from $E[Y_{ij}|A_i=a]$ under informative cluster-population sizes \citep{wang2022two}.} Once the weight is properly scaled, all weights will lead to consistent estimators for $\Delta$. Particularly, the exchangeable correlation structure $\bw(N_i) =  \textup{Corr}(\rho)^{-1} \bone_{N_i}$ will lead to the same estimator as independence correlation structure $\bw(N_i) = \bone_{N_i}$ since $\textup{Corr}(\rho)^{-1} \bone_{N_i} = \{1+(N_i-1)\rho\}^{-1}\bone_{N_i}$ is proportional to $\bone_{N_i}$. In addition to this insight, we provide the optimal weights in the following result.

\begin{proposition}\label{prop:1}
    Given the same conditions as in Theorem 1,  the optimal weights $\bw(N_i)$ that minimize the asymptotic variance of $\widehat{\Delta}$ is 
    \begin{equation}\label{eq:opt-weights}
     \bw(N_i)^{\textup{opt}} = c(N_i) E\left[IF(\bO_i) IF(\bO_i)^\top\big| N_i\right]^{-1} \bone_{N_i},   
    \end{equation}
    where $IF(\bO_i)  \in \mathbb{R}^{N_i}$ is the unweighted vector of individual influence function defined in Supplementary Material C.1 and $c(N_i)>0$ is an arbitrary scalar.
\end{proposition}

Proposition~\ref{prop:1} sheds light on the recipe for improving the precision of the doubly-robust estimator by adjusting for the working correlation structure. To obtain the optimal weights, one needs to compute the matrix inverse in Equation~(\ref{eq:opt-weights}) using observed data, which can be computationally unstable when we have a limited number of clusters. 
However, under an important special case, $\bw(N_i)^{\textup{opt}}$ has a simple form.  We assume that the individual-level data in a cluster are exchangeable; that is, any permutation of their indices $j$ does not change their joint distribution.
In practice, this is often the default setting used in trial planning \citep{turner2017review}. 
In this case, direct calculation shows that $\bw(N_i)^{\textup{opt}}\propto\bone_{N_i}$, indicating that the constant weight leads to optimal efficiency. Due to its optimality, we recommend using the simplest independence working correlation structure in practice, especially when the number of clusters is small and $N_i$ has large variability.

\subsubsection{Estimating the nuisance functions using machine learning}\label{subsc:ml}
By the semiparametric efficiency theory, the variance of $\widehat{\Delta}$ is minimized when the nuisance functions $\kappa$ and $\eta$ are both correctly specified (The formal statement is provided by Proposition A in Supplementary Material C.2). To optimize the precision of $\widehat{\Delta}$, we incorporate debiased machine learning with cross-fitting \citep{chernozhukov2018double}. Compared to parametric models considered in Equation~\eqref{eq: est}, machine learning models can better approximate the true nuisance functions in a larger and more complex functional model space, thereby achieving optimal precision.

This procedure has the following steps. 
First, we randomly partition $m$ clusters into $K$ parts such that the numbers of clusters across parts differ by at most 1. 
We denote the cluster indices for part $k$ as $\mathcal{Q}_k$ and the rest as $\mathcal{Q}_{-k}$. Next, for each $k =1,\dots, K$, we  use data $\mathcal{Q}_{-k}$ to train machine learning models $\widehat{\kappa}^{(k)}(A_i,{\bfB_i})$ and $\widehat{\eta}^{(k)}(A_i,{\bfB_i})$ for $P(R_{ij}^Y=1|A_i, \bfB_{i})$ and $E[Y_{ij}|R_{ij}=1,A_i, \bfB_{i}]$, respectively, and evaluate the trained models on  $\mathcal{Q}_k$.
Finally, we denote the nuisance function estimators as $\widehat{\kappa}(A_i,{\bfB_i}) = \sum_{k=1}^K \mathbb{I}\{i\in \mathcal{Q}_k\} \widehat{\kappa}^{(k)}(A_i,{\bfB_i})$ and $\widehat{\eta}(A_i,{\bfB_i})= \sum_{k=1}^K \mathbb{I}\{i\in \mathcal{Q}_k\} \widehat{\eta}^{(k)}(A_i,{\bfB_i})$, where $\mathbb{I}$ is the indicator function. 
For the machine learning models, we can choose any method such that the following condition holds.

\begin{assumption}[Machine learning models]\label{assp:5}
(1) $\widehat{\kappa}(A_i,{\bfB_i}) - P(R_{ij}^Y=1|A_i, \bfB_{i}) = o_p(m^{-1/4})$ and $\widehat{\eta}(A_i,{\bfB_i}) - E[Y_{ij}|R_{ij}=1,A_i, \bfB_{i}] = o_p(m^{-1/4})$ in $L_2$-norm. (2) $\widehat{\kappa}(A_i,{\bfB_i})$ is uniformly bounded in $(0,1)$. (3) $N_i$ is upper bounded by some large constant.  
\end{assumption}

Assumption \ref{assp:5}(1) places constraints on the convergence rate of the nuisance function estimators to the truth to be faster than $m^{-1/4}$. This condition is much weaker than the classical  $m^{-1/2}$ rate and can be achieved by many machine learning algorithms, such as deep neural networks \citep{farrell2021deep} and highly-adaptive lasso \citep{benkeser2016highly}. For  clustered data, we can simplify the estimator by setting $\widehat{\kappa}(A_i,
\bfB_i) = \widehat{\kappa}(A_i,\bB_{ij}, h(\bfB_i))$ for a summary, potentially high-dimensional function $h$.
In addition, we assume Assumptions \ref{assp:5}(2) and \ref{assp:5}(3) to control the remainder terms and allow for variance estimation. 

With the new nuisance function estimators, we update Equation~(\ref{eq: est}) correspondingly and estimate $\widehat{\Delta}^{\textup{ml}}$ in the same way.  Its variance estimator,  $\widehat{Var}(\widehat{\Delta}^{\textup{ml}})$, is provided in Supplementary Material A.
Theorem~\ref{thm2} below characterizes the asymptotic behavior of $\widehat{\Delta}^{\textup{ml}}$.

\begin{theorem}\label{thm2}
    Given Assumptions \ref{assp1}-\ref{assp:5}, $\{\widehat{Var}(\widehat{\Delta}^{\textup{ml}})\}^{-1/2} (\widehat{\Delta}^{\textup{ml}} - \Delta) \xrightarrow{d} N(0,1)$, and\\ $m \widehat{Var}(\widehat{\Delta}^{\textup{ml}})$ converges in probability to $v^{opt}$, where $v^{opt}$ is the asymptotic variance of $\widehat{\Delta}$ in Theorem \ref{thm1} when the parametric nuisance functions are correctly specified. 
\end{theorem}

Theorem~\ref{thm2} differs from Theorem~\ref{thm1} in that it achieves consistency and the optimal variance (with respect to $\widehat{\Delta}$), leveraging the flexibility of machine learning.
{Given a moderate to
large number of clusters, $\widehat{\Delta}^{\textup{ml}}$ can often lead to high precision. In contrast, with fewer clusters, parsimonious parametric models may offer more stable performance, as machine learning may overfit. Their comparisons are demonstrated through simulations in Section~\ref{sec: simulation}.}


\begin{remark}
In Supplementary Material C, we prove that the asymptotic variance $v^{opt}$ with $w_j(N_i)=1$ achieves the semiparametric efficiency lower bound when $\{(Y_{ij}, R_{ij}):j=1,\dots, N_i\}$ are conditionally independent given $(A_i, \bfB_i)$. In this special case, all intracluster correlations are captured by observed covariates, and individual distributions are exchangeable within a cluster, leading to the optimal statistical efficiency of $\widehat{\Delta}^{\textup{ml}}$ with constant weights. 
\end{remark}

\subsubsection{Modeling the propensity score for treatment assignment}
Modeling the propensity score for treatment assignment has been established as a technique to improve precision in randomized trials (for example, \citealp{balzer2016targeted}).
However, the extent of its additional benefit to doubly robust estimation in CRTs is unclear.
In Supplementary Material C.4, we describe the steps to include this model, and provide a theoretical guarantee that it \textit{never increases} the asymptotic variance of $\widehat{\Delta}$ when outcome missingness is correctly modeled. This result justifies the benefits of this technique in CRTs in improving precision.

\section{Estimation under uniform within-cluster sampling}\label{sec: est-sampling}
When a cluster only enrolls a subset of individuals into the CRT, we have $M_i < N_i$ and $N_i$ is subject to missingness. 
However, since neither outcomes nor individual-level covariates are observed for non-participants, the enrollment mechanism is unidentifiable from observed data. In this case, we provide sufficient conditions, Assumption~\ref{assp:6} in place of Assumption~\ref{assp3}, to characterize a basic scenario for doubly-robust estimation.

\begin{assumption}[rMAR with uniform sampling]\label{assp:6}
(1) For each $i=1,\dots, m$, the enrollment vector $\bE_i = (E_{i1},\dots, E_{iN_i})$ is independently generated by $P(\bE_i = \boldsymbol{e}|M_i, N_i) = {\binom{N_i}{M_i}}^{-1}$ for any $N_i$-dimensional binary vector $ \boldsymbol{e}$ with $M_i$ ones. (2) Given observed cluster-level covariates $(C_{i1}^*, \dots, C_{iq}^*, M_i)$ and $R_i^N = 0$, $N_i$ is conditionally independent of $(R_{ij}^Y, Y_{ij}, X_{ij1}^*,\dots, X_{ijp}^*)$. (3) $R_{ij}^Y$ is conditionally independent of $Y_{ij}$ given $(E_{ij}=1,A_i, \bB_{ij})$ for each $j$ with $E_{ij}=1$.
\end{assumption}

Assumption ~\ref{assp:6}(1) characterizes the uniform distribution of $\bE_i$ and rules out selection bias. 
Assumption \ref{assp:6}(2) handles missing $N_i$ by assuming non-informative cluster-population sizes. In other words, cluster-level covariates explain all the correlations between $N_i$ and individual-level data whenever $N_i$ is unobserved. 
Finally, Assumption \ref{assp:6}(3) updates Assumption~\ref{assp3} for missing outcomes. Here, rMAR is still assumed, while the conditioning variables are restricted to $E_{ij} =1$ (such that the data for individual $j$ can be observed), $A_i$, and $\bB_{ij}$ (instead of $\bfB_i$ in Assumption~\ref{assp3}, which involves covariate data from non-participants). 
Altogether, Assumption~\ref{assp:6} provides a scenario where the trial participants are representative samples of the entire cluster and missing data are ignorable given observed data. 

Under this setting, we specify the parametric nuisance function estimators as $\eta(A_i, \bB_{ij};\btheta_Y)$ and $\kappa(A_i, \bB_{ij};\btheta_R)$ and compute $\widehat{\btheta}_Y$ and $\widehat{\btheta}_R$ in the same way as in Section \ref{sec: dr-estimation}.
We next define
\begin{equation*}
    D_{ij}^\dagger(a, \btheta_Y, \btheta_R) = \frac{E_{ij}}{M_i}\left[ \frac{I\{A_i=a\}}{\pi^a(1-\pi)^{1-a}}     \frac{R_{ij}^Y \{Y_{ij} - \eta(a, \bB_{ij};\btheta_Y)\}}{ \kappa(a, \bB_{ij};\btheta_R)} + \eta(a, \bB_{ij};\btheta_Y)\right]
\end{equation*}
for $j=1,\dots, N_i$ and $a = 0,1$. The estimator for $\Delta$ is then computed as $\widehat{\Delta}^\dagger = f(\widehat\mu_1^\dagger, \widehat\mu_0^\dagger)$, where $\widehat{\mu}_a^\dagger = m^{-1} \sum_{i=1}^m \sum_{j=1}^{N_i} D_{ij}^\dagger(a, \widehat{\btheta}_Y, \widehat{\btheta}_R)$ for $a = 0,1$. 
The variance estimator $\widehat{Var}(\widehat{\Delta}^\dagger)$ also uses the sandwich variance estimation as in Section \ref{sec: dr-estimation}. Compared to $\widehat{\Delta}$, the major difference of $\widehat{\Delta}^\dagger$ is the constant weight, $M_i^{-1}E_{ij}$, among sampled individuals.
Theorem~\ref{thm3} below provides parallel results of double-robustness and asymptotic normality for $\widehat{\Delta}^\dagger$ to Theorem~\ref{thm1}.

\begin{theorem} \label{thm3}
    Given Assumptions \ref{assp1}, \ref{assp2}, \ref{assp:6} and regularity conditions in Supplementary Material B, if either $\kappa(A_i, \bB_{ij};\btheta_R^*) = P(R_{ij}^Y=1|E_{ij}=1, A_i, \bB_{ij})$  or $\eta(A_i, \bB_{ij};\btheta_Y^*)=E[Y_{ij}|R_{ij}^Y = 1, E_{ij}=1, A_i,\bB_{ij}]$, then $\{\widehat{Var}(\widehat{\Delta}^\dagger)\}^{-1/2} (\widehat{\Delta}^\dagger - \Delta) \xrightarrow{d} N(0,1)$.
\end{theorem}

Beyond parametric models, machine learning can be incorporated into $\widehat{\Delta}^\dagger$ in the same way as $\widehat{\Delta}^{\textup{ml}}$, and the same efficiency comparison applies. 


\section{Sensitivity analyses for missing data assumptions}\label{sec: sensitivity}
To estimate $\Delta$, we have made structural assumptions to deal with the sampling procedure and missingness in outcomes. However, these assumptions may not hold under selection bias and outcome MNAR.
To evaluate such effects, we study the bias of our proposed estimators without putting any constraint on the sampling or missing data mechanism. To present a relatively clean result, we assume $N_i$ is observed and $E_{ij}$ is a pre-randomization variable included in $\bB_{ij}$. Proposition~\ref{prop:5} quantifies the asymptotic bias of the doubly-robust estimator.

\begin{proposition}\label{prop:5}
Assume Assumptions~\ref{assp1}, \ref{assp2}, and regularity conditions in Supplementary Materials B. If either nuisance model $\kappa$  or $\eta$ is correctly specified, then $\widehat{\mu}_a^\dagger \xrightarrow{P}\underline{\mu}_a$, where
    \begin{equation}\label{eq: sensitivity-bias}
        \underline{\mu}_a - E[Y_{ij}|A_{i}=a] = E\left[\frac{N_i-M_i}{N_i}\delta_{a}(M_i, N_i)\right] + E[E[(1-R_{ij}^Y)\gamma_{a}(\bB_{ij})|E_{ij}=1,A_i=a, M_i, N_i]],
    \end{equation}  
    where $\delta_{a}(M_i, N_i) = E[Y_{ij}| E_{ij}=1, A_i=a, M_i, N_i] - E[Y_{ij}|E_{ij}=0, A_i=a, M_i, N_i]$ and $\gamma_{a}(\bB_{ij}) =E[Y_{ij}|R_{ij}^Y=1,E_{ij}=1,A_i=a, \bB_{ij}] - E[Y_{ij}|R_{ij}^Y=0,E_{ij}=1, A_i=a, \bB_{ij}].$
\end{proposition}

In Equation~(\ref{eq: sensitivity-bias}), the bias of $\widehat{\mu}_a^\dagger$ has two components, selection bias from within-cluster sampling and the bias from outcome MNAR. The first component, $E\left[\frac{N_i-M_i}{N_i}\delta_{a}(M_i, N_i)\right]$, is determined by the outcome mean difference between participants and non-participants, $\delta_{a}(M_i, N_i)$, and the proportion of non-participants, $\frac{N_i-M_i}{N_i}$. This source of bias increases if either $\delta_{a}(M_i, N_i)$ or $\frac{N_i-M_i}{N_i}$ becomes larger. If $N_i=M_i$ as considered in Section \ref{sec: dr-estimation}, this component of bias is zero. Similarly,  the second component, $E[E[(1-R_{ij}^Y)\gamma_{a}(\bB_{ij})|E_{ij}=1,A_i=a, M_i, N_i]]$, is determined by the conditional mean difference between missing outcomes and observed outcomes among participants, $\gamma_{a}(\bB_{ij})$, and the conditional proportion of missing outcomes,  $P(R_{ij}^Y=0|E_{ij}=1,A_i=a,\bB_{ij})$.  When more outcomes are missing, and missing outcomes have larger difference from observed outcomes, the bias of $\underline{\mu}_a$ becomes larger. 

While Equation~\eqref{eq: sensitivity-bias} involves inestimable quantities, it can still be used to perform a sensitivity analysis. The sensitivity analysis aims to identify the tipping point, i.e., increasing $|\delta_a(M_i,N_i)|$ and $|\gamma_{a}(\bB_{ij})|$ such that $\widehat{\Delta}$ is no longer statistically significant. If $\widehat{\Delta}$ is insignificant in the original analysis, the sensitivity analysis can visualize the impact of missing data on bias and provide qualitative evidence. The larger the tipping point is, the less sensitive our statistical inference is to violations of missing data assumptions. 
In practice, one can set $\delta_a(M_i,N_i) \equiv \delta_a$ and $\gamma_{a}(\bB_{ij}) \equiv \gamma_a$ to find constant tipping points. As an example, when $\Delta = E[Y_{ij}|A_i=1] - E[Y_{ij}|A_i=0]$, then the asymptotic bias of $\widehat{\Delta}^\dagger$ is
$$E\left[\frac{N_i-M_i}{N_i}\right](\delta_1-\delta_0) + E[h_1(M_i,N_i)]\gamma_1 - E[h_0(M_i,N_i)]\gamma_0,$$
where $E\left[\frac{N_i-M_i}{N_i}\right]$ and $h_a(M_i,N_i) = P(R_{ij}^Y=0|A_i=a,E_{ij}=1, M_i, N_i)$ are estimable from the observed data. Then a grid search of tipping points on $(\delta_1-\delta_0, \gamma_1, \gamma_0)$ can reflect the sensitivity of our estimator. Finally, even when $N_i$ is unobserved, we can use bounds for $E\left[\frac{N_i-M_i}{N_i}\right]$ and approximations of $h_a(M_i,N_i)$ by $P(R_{ij}^Y=0|A_i=a,E_{ij}=1)$ as a practical solution to perform the sensitivity analysis. In the special case with $N_i=1$, this approach reduces to the pattern mixture model for individually-randomized trials \citep{little1993pattern}.

\section{Simulation study}\label{sec: simulation}
\subsection{Simulation design}
We conduct two simulation studies; the first one supports our theoretical results in Sections~\ref{sec: dr-estimation} and \ref{sec: est-sampling} and the second one demonstrates the sensitivity method in Section~\ref{sec: sensitivity}.

In the first simulation study, we consider the combination of eight scenarios defined by (i) no sampling (Assumption~\ref{assp3}) versus uniform sampling with missing $N_i$ (Assumption~\ref{assp:6}), (ii) a small ($m=30$) versus large ($m=100$) number of clusters, and (iii) a small ($p_m=10\%$) versus moderate ($p_m=30\%$) proportion of missing  outcomes and covariates. 
For each scenario, we independently generate data for each cluster $i$ as described in~Table \ref{tab:DGP}. 
Across all eight scenarios, the covariate missingness mechanism involves MNAR, while outcomes follow the rMAR assumption. We generate 1000 data replicates under each scenario to evaluate the performance of the estimators. For each data replicate, we compute five estimators detailed in Table~\ref{tab:estimators}.  {Across all estimators that involve modeling covariates, the outcome missingness model is correctly specified, while parametric outcome regression models are misspecified.}

\begin{table}[]
    \centering
    \caption{Data generating distributions for simulated data. A Normal distribution with mean $\mu$ and variance $\sigma^2$ is denoted as $N(\mu,\sigma^2)$. A Bernoulli distribution with success probability $p$ is denoted as $\mathcal{B}(p)$.}
    \label{tab:DGP}
    \begin{tabular}{p{6.5cm}p{10cm}}
    \hline
    Random variables     &  Distributions\\
    \hline
    Cluster-population size  $N_i$   & A uniform distribution on integers between 10 and 90, {where the coefficient of variation is 0.47.}\vspace{0.1in}\\
    Cluster-level covariate $C_i $&  $C_i \sim N(N_i/10, 1)$ \vspace{0.1in}\\
    Individual-level covariates $X_{ij1}, X_{ij2}$& $X_{ij1} \sim \mathcal{B}(N_i/100)$, $X_{ij2} = b_{ij2} + I\{C_i>0\}c_i$, with $b_{ij2} \sim N(C_i\overline{X}_{i1},1)$, $c_i \sim N(0,1)$, and $\overline{X}_{i1} = \sum_{j=1}^{N_i}X_{ij1}/N_i$ \vspace{0.1in}\\
    Covariate non-missing indicators $R_i^C, R_{ij1}^X, R_{ij2}^X$ &  $ R_i^C \sim \mathcal{B}(\textup{expit}\{\textup{logit}(1-p_m)+(C_i-N_i/10)/2\}) $,\,\,\,\,\, $ R_{ij1}^X \sim \mathcal{B}(1-p_m) $, $ R_{ij2}^X \sim \mathcal{B}(\textup{expit}\{\textup{logit}(1-p_m)+(C_i-N_i/10)/2\})$ \vspace{0.1in}\\
     Treatment $A_i$   & $A_i \sim \mathcal{B}(0.5)$ \vspace{0.1in}\\
     Outcome $Y_{ij}$ & $ Y_{ij} = 0.25\sin(C_i) e^{X_{ij1}} |X_{ij2}+1| + 10A_iX_{ij1} + \gamma_i + \varepsilon_{ij}$ with $\gamma_i \sim N(0,1)$ as a cluster-level random intercept and $\varepsilon_{ij} \sim N(0,1)$ as the random error. \vspace{0.1in}\\
     Outcome non-missing indicator $R_{ij}^Y$ & $R_{ij}^Y \sim \mathcal{B}(\textup{expit}\{\textup{logit}(0.99-0.2p_m)-(1.5+5p_m)R_{ij1}^XX_{ij1}\})$ \vspace{0.1in}\\
     With-cluster sampling & The cluster size $M_i$ uniformly sampled from integers between $N_i/2-3$ and $N_i/2+2$, and $M_i$ out of $N_i$ individuals from cluster $i$ randomly sampled and labeled as $E_{ij}=1$ \\
     Final observed data for cluster $i$ &$\bfO_i = \{(R_{ij}^Y, R_{ij}^Y Y_{ij}, A_i, X_{ij1}^*, X_{ij2}^*, C_i^*, N_i^*, M_i):j=1,\dots, N_i;E_{ij}=1\}$ \\
    \hline
    \end{tabular}
\end{table}

\begin{table}[]
    \centering
    \caption{Estimators in the simulation study. All estimators that involve modeling covariates use the same set of covariates. }
    \label{tab:estimators}
    \begin{tabular}{p{2cm}p{14cm}}
    \hline
    Estimator & Implementation\\
    \hline
    Unadjusted     &  \vspace{-0.2in}\begin{itemize}
        \item The mean difference of
cluster-average observed outcomes
\item No adjustment for missing data
    \end{itemize} \\
    IPW     &    \vspace{-0.2in}\begin{itemize}
        \item The Inverse Probability Weighted GEE defined in Section 2.2 of \cite{prague2016accounting}
        \item Correctly specified logistic regression for outcome missingness 
        \item no outcome regression model
\item Independence correlation structure
    \end{itemize}    \\
    DR & \vspace{-0.2in}\begin{itemize}
        \item The doubly-robust estimator proposed by \cite{prague2016accounting}
        \item IPW + (misspecified) outcome linear regression models on main terms
    \end{itemize} \\
    DR-PM & \vspace{-0.2in}\begin{itemize}
        \item The proposed estimator in Section~\ref{sec: dr-estimation} with parametric nuisance models
        \item Same outcome missing model and outcome regression model as DR, with an additional logistic regression for treatment assignment
        \item Constant weights $\bw(N_i)$
    \end{itemize} \\
    DR-ML & \vspace{-0.2in}\begin{itemize}
        \item The proposed estimator in Section~\ref{subsc:ml}
        \item An ensemble model of generalized linear models, regression trees, and generalized additive models for outcome missingness and observed outcomes.
        \item Constant weights $\bw(N_i)$
    \end{itemize} \\
    \hline
    \end{tabular}
\end{table}

In the second simulation study, we generate the observed data and compute the proposed estimators (DR-PM, DR-ML) in the same way as in the first simulation study. Then, we use Equation~\eqref{eq: sensitivity-bias} to search the tipping points at the 0.05 level. For demonstration, we only search for constant tipping points, which assumes $\delta_a(M_i,N_i) \equiv \delta_a$ and $\gamma_a(\bB_{ij}) \equiv \gamma_a$. Since the data-generating distribution of $R_{ij}^Y$ does not depend on $A_i$, 
the bias of $\widehat{\Delta}^\dagger$ is a function of $(\delta_1-\delta_0, \gamma_1-\gamma_0)$. Fixing $\delta_1 - \delta_0 \in\{0,1,2,3,4\}$, the tipping point is the minimum value of $\gamma_1 - \gamma_0$ such that the bias-corrected confidence interval covers zero.


\subsection{Simulation results}
Table~\ref{tab:sim1} summarizes the results of the first simulation study. Across all eight scenarios, the proposed estimators have negligible bias and nominal coverage probabilities. {This is because DR-PM is doubly-robust and correctly specifies the outcome missingness model; and DR-ML well approximated both nuisance models with the ensemble learner.} Comparing DR-PM with DR-ML,  machine learning further reduces the variance by $29\%-47\%$, demonstrating its precision gain in analyzing CRTs. {However, a potential caveat is that variance estimation for DR-ML can be anti-conservative in finite samples, underestimating the variance by 10\%-20\% under our data-generating distribution and leading to 1\%-3\% under-coverage of the confidence intervals. These findings continue to hold under the null of no treatment effect or when $m$ is as few as $10$; additional results are provided in Supplementary Material E.}

Unlike the proposed estimators, the three comparison methods all have bias and under-coverage. Since the unadjusted estimator completely ignores the missing data, {bias ranges from $-0.36$ to $-1$, causing small to moderate inflated type I error rates across scenarios.} 
While the IPW and DR estimators correctly model the missing outcome mechanism, they target a different estimand due to informative cluster sizes, leading to a bias of 1.09. {Because the DR estimator is more efficient than the IPW estimator, the same magnitude of bias leads to more severe type I error rate inflation for the former.}



\begin{figure}[htbp]
    \centering
    \includegraphics[width=0.9\textwidth]{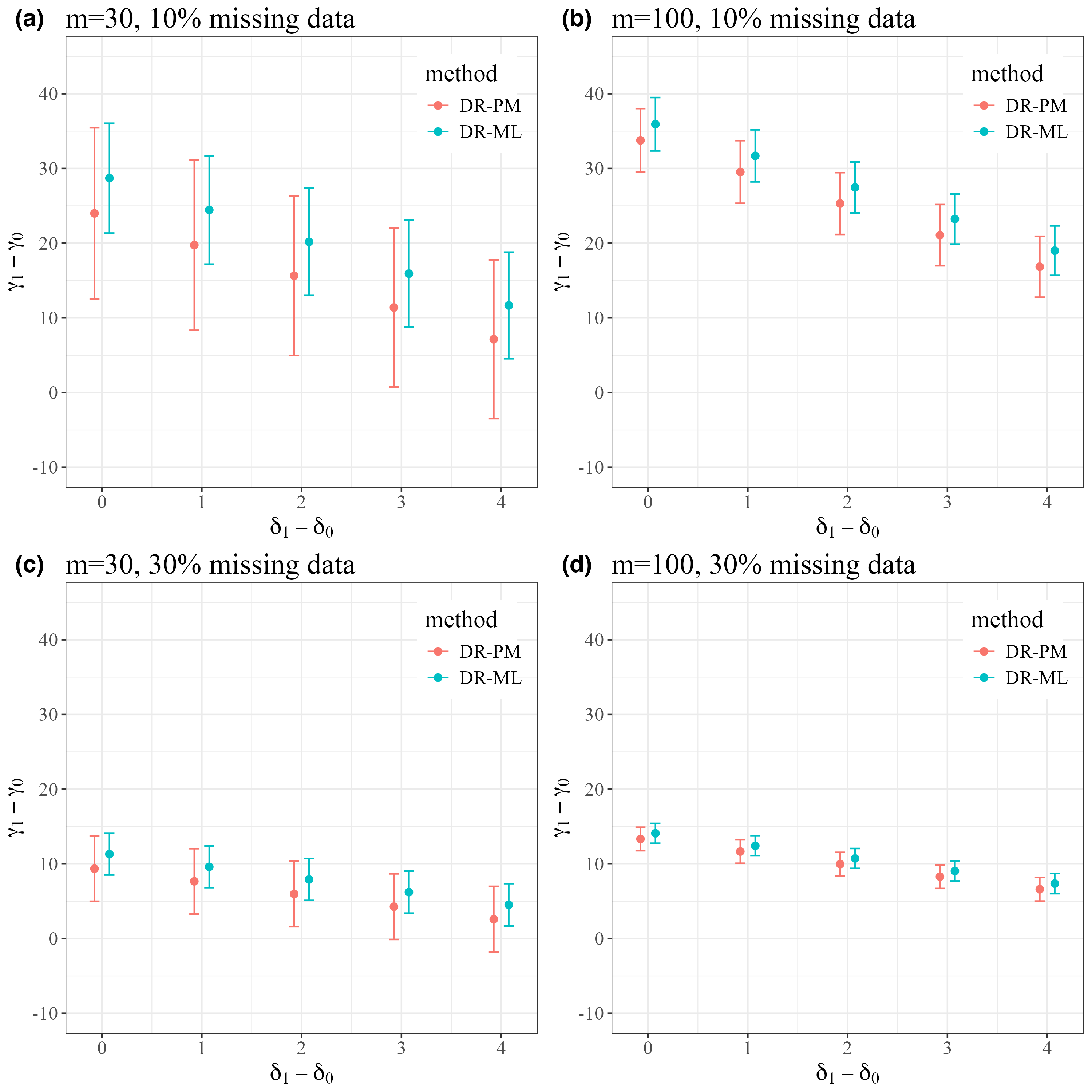}
    \caption{Results for the second simulation study. The x-axis $\delta_1 - \delta_0$ measures the selection bias on the average treatment effect from non-uniform sampling, and {the y-axis $\gamma_1 - \gamma_0$ measures the minimum required difference in the average treatment effect between observed and missing outcomes to overturn the study findings, i.e., making the confidence interval covers zero. Fixing $\delta_1 - \delta_0$, larger $\gamma_1 - \gamma_0$ represents more robustness to violations of missing data assumptions.} In each plot, the dots with error bars represent the mean value plus/minus the standard error obtained from 1000 simulations. }
    \label{fig:sim2}
\end{figure}

Figure~\ref{fig:sim2} presents the result of the second simulation study. In the absence of selection bias (i.e., $\delta_1 -\delta_0 =0$), the proposed methods are less sensitive to missing data assumptions when there is less missing data or a larger number of clusters. 
Intuitively, less missing data reduces the influence of the unobserved data distribution, while a larger number of clusters decreases the variance, pushing the confidence interval boundaries further from zero. 
We also observe that the DR-ML estimator is less sensitive than the DR-PM estimator. As $\delta_1 -\delta_0$ increases, selection bias contributes more substantially to estimation bias, resulting in greater sensitivity of the estimates to the missing data assumptions.


\begin{table}[htbp]
    \centering
    \caption{Results for the first simulation study. ESE: empirical standard error. ASE: average of standard error estimators. CP: coverage probability of 95\% confidence intervals based on $t$-distributions with degree of freedom $m-q$, where $q$ is the number of covariates used in the analysis. }
    \label{tab:sim1}
    \begin{tabular}{ccrrrrrcrrrr}
    \hline
          \multirow{2}{*}{Sampling}  &\multirow{2}{*}{$m$} & \multirow{2}{*}{Estimator}&   \multicolumn{4}{c}{10\% missing data}& &\multicolumn{4}{c}{30\% missing data}\\
          \cline{4-7} \cline{9-12}
           & & & 
     Bias& ESE& ASE& CP&& Bias& ESE& ASE& CP\\
    \hline
  \multirow{10}{*}{No}&\multirow{5}{*}{30}& Unadjusted& -0.36 & 1.15 & 1.11 & 0.92&& -1.00 & 1.07 & 1.08 & 0.83 \\ 
  && IPW& 1.02 & 1.40 & 1.26 & 0.86&& 1.03 & 1.34 & 1.27 & 0.88 \\ 
  && DR & 1.06 & 1.18 & 1.05 & 0.84&& 1.05 & 1.18 & 1.05 & 0.84 \\
  && DR-PM& -0.00 & 1.13 & 1.08 & 0.94&& -0.02 & 1.14 & 1.08 & 0.92 \\
  && DR-ML& 0.00 & 0.82 & 0.77 & 0.93&& -0.01 & 0.89 & 0.80 & 0.92 \\ 
  \cline{2-12}
  &\multirow{5}{*}{100}& Unadjusted& -0.34 & 0.58 & 0.61 & 0.92&& -0.98 & 0.57 & 0.59 & 0.61 \\ 
  && IPW& 1.07 & 0.71 & 0.72 & 0.70&& 1.10 & 0.72 & 0.72 & 0.69 \\ 
  && DR & 1.07 & 0.58 & 0.57 & 0.56&& 1.11 & 0.59 & 0.58 & 0.54 \\ 
  && DR-PM& -0.01 & 0.49 & 0.51 & 0.96&& 0.02 & 0.50 & 0.52 & 0.96 \\ 
  && DR-ML& -0.02 & 0.40 & 0.37 & 0.92&& -0.00 & 0.42 & 0.40 & 0.94 \\ 
  \hline
  \multirow{10}{*}{Yes}&\multirow{5}{*}{30}& Unadjusted& -0.36 & 1.17 & 1.13 & 0.93&& -0.99 & 1.15 & 1.10 & 0.83 \\ 
  && IPW& 1.06 & 1.41 & 1.27 & 0.86&&  1.05 & 1.43 & 1.30 & 0.85 \\
  && DR & 1.08 & 1.20 & 1.07 & 0.85&& 1.06 & 1.20 & 1.08 & 0.84 \\
  && DR-PM& 0.05 & 1.13 & 1.07 & 0.95&& -0.02 & 1.16 & 1.17 & 0.95 \\ 
  && DR-ML& 0.02 & 0.86 & 0.79 & 0.94&& -0.02 & 0.92 & 0.83 & 0.92 \\ 
  \cline{2-12}
  &\multirow{5}{*}{100}& Unadjusted& -0.36 & 0.61 & 0.61 & 0.90&& -1.00 & 0.58 & 0.60 & 0.63 \\ 
  && IPW& 1.06 & 0.73 & 0.73 & 0.72 && 1.09 & 0.75 & 0.74 & 0.70 \\ 
  && DR & 1.07 & 0.60 & 0.58 & 0.57&& 1.11 & 0.62 & 0.59 & 0.53 \\
  && DR-PM& -0.01 & 0.50 & 0.52 & 0.96&& 0.00 & 0.52 & 0.54 & 0.96 \\ 
  && DR-ML& -0.03 & 0.40 & 0.38 & 0.94 && -0.02 & 0.44 & 0.41 & 0.93 \\ 
    \hline
    \end{tabular}
\end{table}

\vspace{-20pt}

\section{Data application}\label{sec: data-application}
The Work, Family, and Health Study \citep{WFHS} is a CRT for improving the well-being of employees and reducing their work-family conflict. In this trial, 56 working groups (clusters) in an anonymized company were randomized in a 1:1 ratio to receive a workplace intervention (treatment) or not (control). Each cluster has 7-60 employees (mean 24, standard error 12), while participants of this CRT range 3-42 per cluster (mean 15, standard error 8). The outcome we focus on is control over work hours at the 6-month follow-up, a continuous measure ranging from 1 to 5, and has 20\% missing data. We include two covariates: the baseline value of the outcome measure (13\% missing data), and group job functions (core or supporting) with no missingness. 
We consider estimating the average treatment effect defined in \eqref{eq:estimand} using the five estimators compared in Section~\ref{sec: simulation}.

Table~\ref{tab: data-application} summarizes the results of our data application. Across all five estimators, their point estimates are relatively close, and their confidence intervals all exclude zero. 
Because the true data-generating mechanism is unknown and no data replications are available, the observed similarities among estimators may be due to missingness completely at random.
{Finally, a sensitivity analysis (detailed in Supplementary Material F) indicates that the results are relatively robust to the missing data assumptions.}
\begin{table}[htbp]
\centering
\caption{Results of re-analysis of the WFHS data.}\label{tab: data-application}
\begin{tabular}{rrrrr}
  \hline
 & Estimate & Standard error & 95\% confidence interval \\ 
  \hline
Unadj & 0.17 & 0.09 & (0.00, 0.34) \\ 
  IPW & 0.18 & 0.07 & (0.03, 0.32) \\ 
  DR  & 0.22 & 0.05 & (0.12, 0.32) \\ 
  DR-PM & 0.20 & 0.06 & (0.07, 0.32) \\ 
  DR-ML & 0.20 & 0.05 & (0.09, 0.31) \\ 
   \hline
\end{tabular}
\end{table}


\vspace{-0.2in}
\section{Discussion}\label{sec: discussion}

Building on the literature of doubly robust estimation, we have developed novel approaches to address missing data across multiple domains while maximizing precision and rigorously assessing sensitivity to missing data assumptions. 
In the sensitivity analysis, the tipping points provides an interpretable metric for the robustness to violations of missing data assumptions, offering researchers a principled way to evaluate the reliability of their findings. 

{To address missing outcomes in CRTs, \citet{prague2016accounting} proposed a doubly robust estimator under the assumption of non-informative cluster sizes. Our proposed methods build on theirs but differ in several key aspects.
First, we account for potential bias from informative cluster sizes by scaling the weights $w_j(N_i)$ derived from the working correlation structure. Second, our method accommodates missing data across multiple domains including covariate missingness and non-participation.
Third, we enhance efficiency through three methodological insights, whose validity is demonstrated by simulation studies. Finally, we develop a sensitivity analysis framework tailored to our missing data assumptions.
}

{An alternative approach in this space is the two-stage TMLE framework \citep{balzer2023two, nugent2024blurring, benitez2021comparative}, which employs doubly robust estimation for missing outcomes within each cluster, followed by cluster-level analyses. This method maintains double robustness under two key conditions that are different from our work: (1) within-cluster correlations are sufficiently weak to apply the central limit theorem within clusters, and (2) cluster sizes grow faster than the number of clusters. In Supplementary Material E, we present simulation studies comparing our methods with two-stage TMLE approaches. Notably, \cite{nugent2024blurring} proposed methods to handle missing covariates and subsampling under MAR, where both missingness mechanisms can be directly estimated from observed data. Our framework, in contrast, addresses a distinct setting with covariates MNAR and within-cluster sampling, where non-participants have no observed data. This distinction highlights the applicability of our approach and sensitivity analysis in scenarios where missingness mechanisms cannot be fully recovered from the observed data.}

{Beyond the simple randomization specified in Assumption 2, our theoretical results also extend to covariate-adaptive randomization schemes, such as stratified randomization on baseline covariates. Following the results of \citet{wang2023model, wang2024asymptotic, bannick2025general}, including the covariates used in the randomization process in the analysis model renders the inference asymptotically equivalent to that under simple randomization. We therefore recommend such covariate adjustment when applying our proposed methods to CRTs with covariate-adaptive designs.}

Under within-cluster sampling, we characterize a basic uniform-sampling scenario where enrollment causes no selection bias. When selection bias may occur, principal stratification \citep{li2022clarifying} with appropriate structural assumptions can identify treatment effects for certain subpopulations, such as individuals who are always enrolled regardless of treatment assignment. While specific covariate-dependent enrollment schemes have been studied by \citet{papadogeorgou2023addressing}, addressing missing data under this more flexible enrollment scheme remains an area for future research. 


\vspace{-0.08in}

\section*{Acknowledgments}
Research in this article was supported by Patient-Centered Outcomes Research Institute Awards\textsuperscript{\textregistered} (PCORI\textsuperscript{\textregistered} Awards ME-2022C2-27676) and the National Institute Of Allergy And Infectious Diseases of the National Institutes of Health (NIH) under Award Number R00AI173395. The statements presented in this article are solely the responsibility of the authors and do not necessarily represent the official views of NIH or PCORI\textsuperscript{\textregistered}, its Board of Governors, or the Methodology Committee. 
We sincerely thank Dr. Laura Balzer for helpful discussions.

\vspace{-0.08in}
\section*{Supplementary Material}
{Web Appendices, Tables, and Figures referenced in Sections \ref{sec: dr-estimation}-\ref{sec: discussion} and code are available with this paper at the Biometrics website on Oxford Academic.}

\vspace{-0.08in}
\section*{Data Availability}
The data used in Section~\ref{sec: data-application} are publicly available at
\url{https://www.icpsr.umich.edu/web/DSDR/studies/36158}.

\bibliographystyle{biom} 
\bibliography{references.bib}





\label{lastpage}

\end{document}


\def\spacingset#1{\renewcommand{\baselinestretch}%
{#1}\small\normalsize} \spacingset{1}

\date{\vspace{-5ex}}

\maketitle

\spacingset{1.5}
\setcounter{page}{1}
\appendix

Section~\ref{asec: variance-estimator} gives the variance estimators for our proposed estimators.
Section~\ref{asec: regularity} contains the regularity conditions for our theoretical results. Section~\ref{asec: efficiency} provides auxiliary theoretical results referred to in Section 3 of the main paper. Section~\ref{asec: proof} contains the proofs. Section~\ref{asec: simulation} provides additional simulation results. Section~\ref{asec: sensitivity} provides the sensitivity analysis for the data application. The R code of our numerical studies is available at \url{https://github.com/BingkaiWang/CRT-missing-data}.

\section{Variance estimators}\label{asec: variance-estimator}
We first provide the variance estimator for $\widehat{\Delta}$ defined in Section 3 of the main paper.
Letting $\btheta = (\Delta, \mu_1, \mu_0, \btheta_Y^\top, \btheta_R^\top)^\top$ denote all parameters involved in the estimation procedure, the estimation of $\btheta$ in the first two steps can be equivalently expressed as $\sum_{i=1}^m \bpsi(\bfO_i;\widehat\btheta) = \bzero$, where the joint estimating function is 
    \begin{align*}
        \bpsi(\bfO_i; \btheta) = \left(\begin{array}{c}
        \Delta - f(\mu_1, \mu_0)\\
        \sum_{j=1}^{N_i}  D_{ij}(a, \btheta_Y, \btheta_R)  -\mu_a,\quad a = 0,1   \\
          \bpsi_R(\bfO_i; \btheta_R)\\
          \bpsi_Y(\bfO_i; \btheta_Y)
        \end{array}\right).
    \end{align*}
The sandwich variance estimator $\widehat{Var}(\widehat{\Delta})$ is computed as the first-row first-column entry of 
    \begin{align*}
        \left\{\sum_{i=1}^m \frac{\partial}{\partial \btheta} \bpsi(\bfO_i;\btheta) \bigg|_{\btheta = \widehat{\btheta}}\right\}^{-1} \left\{\sum_{i=1}^m  \bpsi(\bfO_i;\widehat{\btheta})\bpsi(\bfO_i;\widehat{\btheta})^\top \right\} \left\{\sum_{i=1}^m \frac{\partial}{\partial \btheta} \bpsi(\bfO_i;\btheta) \bigg|_{\btheta = \widehat{\btheta}}\right\}^{-1\ \top}.
    \end{align*}

For the variance estimator of $\widehat{\Delta}^{\textup{ml}}$ defined in Section 3.1.2 of the main paper, we first construct the variance estimator for $\boldsymbol{\mu}^* =(E[Y_{ij}|A_i=1], E[Y_{ij}|A_i=0])$ as
\begin{align*}
\widehat{\bfV}_{\boldsymbol{\mu}}^{\textup{ml}} = \frac{1}{K} \sum_{k=1}^K \widehat{\bfSigma}_k,
\end{align*}
where, denoting $\mathbb{P}_k X = |\mathcal{Q}_k|^{-1} \sum_{i\in \mathcal{Q}_k} X_i$ and $w_j^*(N_i) = (\sum_{j=1}^{N_i} w_j(N_i))^{-1}w_j(N_i)$ as the scaled weight function,
\begin{align*}
       \widehat{\bfSigma}_k &= \mathbb{P}_k  \{\bU(\widehat{\kappa}^{(k)},\widehat{\eta}^{(k)})- \widehat{\boldsymbol{\mu}}_k^{\textup{ml}}\}  \{\bU(\widehat{\kappa}^{(k)},\widehat{\eta}^{(k)})- \widehat{\boldsymbol{\mu}}_k^{\textup{ml}}\}^\top,\\
     \bU(\kappa,\eta) &= (U_1(\kappa,\eta), U_0(\kappa,\eta))^\top, \\
    \widehat{\boldsymbol{\mu}}_k^{\textup{ml}} &= \mathbb{P}_k \bU(\widehat{\kappa}^{(k)},\widehat{\eta}^{(k)}), \\
         U_a(\kappa, \eta) &=  \sum_{j=1}^N w_j^*(N_i) \frac{I\{A=a\}}{\pi^{a}(1-\pi)^{1-a}} \frac{R_{.j}\{Y_{.j} - \eta_{.j}(a)\}}{\kappa_{.j}(a)} + \eta_{.j}(a).
\end{align*}
Then the variance estimator $\widehat{Var}(\widehat{\Delta}^{\textup{ml}})$ for $\widehat{\Delta}^{\textup{ml}}$ is obtained by Delta method based $\widehat{\bfV}_{\boldsymbol{\mu}}^{\textup{ml}}$.

\section{Regularity conditions}\label{asec: regularity}

\begin{assumption}[Regularity conditions for parametric nuisance models]\label{assp:4}
(1) $\btheta \in \mathbf{\Theta}$, a compact set in the Euclidean space. (2) $E[\bpsi(\bfO_i;\btheta)]=\bzero$ has a unique solution $\btheta^* = (\Delta^*,\mu_1^*, \mu_0^*, \btheta_R^*, \btheta_Y^*)$, an interior point of  $\mathbf{\Theta}$. (3) $\bpsi(\bfO_i;\btheta)$ has a finite second moment and second-order derivatives that are uniformly bounded by some integrable function $h(\bfO_i)$ in a small neighborhood of $\btheta^*$. (4) $E[\frac{\partial}{\partial \btheta} \bpsi(\bfO_i;\btheta)|_{\btheta = \btheta^*}]$ exists and is invertible.
\end{assumption}

Assumption~\ref{assp:4} consists of moment and continuity conditions to rule out degenerate conditions for the nuisance functions $\kappa$ and $\eta$, and does not make any parametric assumptions. Of note, Assumption \ref{assp:4}(2) requires a unique root to  $E[\bpsi(\bfO_i;\btheta)]=\bzero$, which is equivalent to the existence of a unique maximizer of the log-likelihood function in large samples if $\bpsi(\bfO_i;\btheta)$ is the corresponding score function. Assumption \ref{assp:4} is satisfied by most commonly used parametric models as long as $\bfO_i$ has finite fourth moments.

\section{Efficiency considerations}\label{asec: efficiency}
\subsection{The influence function defined in Proposition 1}
The influence function $IF(O_i)$ defined in Proposition 1 of the main paper is given as follows.
\begin{align*}
IF(O_i) &= \sum_{a=0}^1 f_a'IF_{i}(a,\bfO_i), \\
IF_i(a,\bfO_i) &= (IF_{i1}(a,\bfO_i), \dots, IF_{iN_i}(a,\bfO_i))^\top,\\
IF_{ij}(a,\bfO_i) &= \frac{I\{A_i=a\}}{\pi^a(1-\pi)^{1-a}}     \frac{R_{ij}^Y \{Y_{ij} - \eta_{ij}(a, \btheta_Y^*)\}}{ \kappa_{ij}(a, \btheta_R^*)} + \eta_{ij}(a, \btheta_Y^*)- \boldsymbol{q}_{1a}^\top \bpsi_R(\bfO_i;\btheta_R^*) - \boldsymbol{q}_{2a}^\top \bpsi_Y(\bfO_i;\btheta_Y^*), \\
\boldsymbol{q}_{1a}^\top &= E\left[\frac{\partial}{\partial \btheta_R} U(a,\bfO_i; \btheta_R,\btheta_Y^*)\Big|_{\btheta_R=\btheta_R^*}\right]^\top E\left[\frac{\partial}{\partial \btheta_R}\bpsi_R(\bfO_i;\btheta_R)\Big|_{\btheta_R=\btheta_R^*}\right]^{-1}, \\
\boldsymbol{q}_{2a}^\top &= E\left[\frac{\partial}{\partial \btheta_Y} U(a,\bfO_i; \btheta_R^*,\btheta_Y)\Big|_{\btheta_Y=\btheta_Y^*}\right]^\top E\left[\frac{\partial}{\partial \btheta_Y}\bpsi_Y(\bfO_i;\btheta_Y)\Big|_{\btheta_Y=\btheta_Y^*}\right]^{-1},\\
U(a,\bfO_i; \btheta_R,\btheta_Y) &= \sum_{j=1}^{N_i} w_j^*(N_i) \left[ \frac{I\{A_i=a\}}{\pi^a(1-\pi)^{1-a}}     \frac{R_{ij}^Y \{Y_{ij} - \eta_{ij}(a, \btheta_Y)\}}{ \kappa_{ij}(a, \btheta_R)} + \eta_{ij}(a, \btheta_Y)\right].
        \end{align*}
\subsection{Optimal variance given correctly specified nuisance functions}\label{subsc:ml}
An important question is how to specify the nuisance function estimators $\kappa(A_i, \bfB_{i};\widehat\btheta_R)$ and $\eta(A_i, \bfB_{i};\widehat\btheta_Y)$ such that the variance of $\widehat{\Delta}$ is minimized. For this purpose, Proposition~\ref{prop:2} shows when correctly specified nuisance function estimators lead to the highest precision.
\begin{proposition}\label{prop:2}
    Given Assumptions 1-4 and pre-specified weights $\bw(N_i)$, if any of the following conditions holds: (1) $P(R_{ij}^Y=1|A_i, \bfB_{i})$ is known, or (2) $P(R_{ij}^Y=1|A_i, \bfB_{i})=\kappa(A_i,\bfB_{i};\btheta_R^*)$ and $\btheta_R^*$ is consistently estimated using independent data source, or (3) $(R_{i1},\dots, R_{iN_i})$ are mutually independent given $(A_i,\bfB_i)$, and 
        $P(R_{ij}^Y=1|A_i, \bfB_{i})=\kappa(A_i,\bfB_{i};\btheta_R^*)$,
    then the estimator $\widehat{\Delta}$ is asymptotically most precise 
    if $\eta(A_i, \bfB_{i};\btheta_Y^*)=E[Y_{ij}|R_{ij}=1,A_i, \bfB_i]$. The optimal asymptotic variance is
    \begin{equation}\label{eq: v-opt}
        v^{opt} = Var\left\{\sum_{a=0}^1\sum_{j=1}^{N_i}  f_a'\frac{w_j(N_i)}{\sum_{j'=1}^{N_i} w_{j'}(N_i)}U_{ij}(a, \bfO_i)\right\},
    \end{equation}
    where $(f_1',f_0')$ is the gradient of $f$ at $(E[Y_{ij}|A_i=1],E[Y_{ij}|A_i=0])$ and
    \begin{align*}
        U_{ij}(a, \bfO_i) = \frac{I\{A_i=a\}}{\pi^a(1-\pi)^{1-a}}     \frac{R_{ij}^Y \{Y_{ij} - E[Y_{ij}|R_{ij}=1,A_i=a, \bfB_i]\}}{ P(R_{ij}^Y=1|A_i=a, \bfB_{i})} + E[Y_{ij}|R_{ij}=1,A_i=a, \bfB_i].
    \end{align*}
\end{proposition}

Proposition~\ref{prop:2} lists three scenarios where, given either known or correctly specified models for the missing outcome propensity score, a correctly specified outcome model $\eta$ yields the most precise estimator asymptotically. Among the three conditions, condition (1) is the most restrictive one since it requires a known mechanism for missing outcomes. In practice, this condition is plausible if outcomes are missing completely at random or if outcome missingness only depends on some discrete variables. Condition (2) is weaker and assumes a correctly specified missing outcome model, but requires the missing outcome model to be estimated using independent data source, such as historical data or cross-fitting (which we introduce later in this subsection). If, otherwise, we use the same data to consistently estimate $\btheta_R^*$ and $\Delta$, then the estimation of $\btheta_R^*$ will result in an extra term (i.e., a projection term to the nuisance tangent space) in the influence function of $\widehat{\Delta}$, as implied by the semiparametric theory \citep{Tsiatis2006}. Because of this extra term, the asymptotic variance of $\widehat{\Delta}$ also involves $E[R_{ij}R_{ij'}|A_i,\bfB_i]$ for $j\ne j'$, which may cause the correctly specified outcome model to be suboptimal under certain data-generating distributions. However, if we consider a special scenario where  $(R_{i1},\dots, R_{iN_i})$ are mutually independent given $(A_i,\bfB_i)$, as stated in condition (3), then the correctly specified outcome model is optimal. Overall, Proposition~\ref{prop:2} is a generalization of \cite{tan2007comment,tsiatis2007comment, cao2009improving} from independent individual data to clustered data in CRTs.

\subsection{Efficiency bounds}
We have derived the variance for the doubly-robust estimator $\widehat{\Delta}$ with machine learning, but is there an estimator, potentially with a different form, that is more precise than $\widehat{\Delta}^{\textup{ml}}$? To answer this question, we need to derive the efficiency bounds for $\Delta$ under our framework. To start with, we present a special case where $\widehat{\Delta}^{\textup{ml}}$ is efficient. 

\begin{proposition}\label{prop:3}
    Given Assumptions~1-3, and assuming that $\{(Y_{ij}, R_{ij}):j=1,\dots, N_i\}$ are mutually independent given $(A_i, \bfB_i)$, then $v^{opt}$ defined in Equation~(\ref{eq: v-opt}) with $w_j(N_i) =1$ is the asymptotic variance lower bound for any regular and asymptotic linear estimator. 
\end{proposition}

Proposition~\ref{prop:3} implies that, when all the intracluster correlations among outcomes and missing outcome indicators are caused by observed covariates, $\widehat{\Delta}^{\textup{ml}}$ (and $\widehat{\Delta}$ with both nuisance function models correctly specified) achieves the efficiency lower bound asymptotically. 

In the general case, however, $v^{opt}$ may not be the efficiency lower bound, but the true lower bound may not be worth pursuing in the CRT setting. 
Since  $\{(Y_{ij}, R_{ij}):j=1,\dots, N_i\}$ can be correlated, achieving the efficiency bound requires modeling $P(\bR_{i}^Y = \boldsymbol{r} |A_i, \bfB_i)$ for any $N_i$-dimensional binary vector $\boldsymbol{r}$ in the support of $\bR_{i}^Y$, and modeling $E[ Y_{ij}| \boldsymbol{r} \circ \bY_i, A_i, \bfB_i]$, i.e., the conditional mean of unobserved outcomes given all observed outcomes, treatment, and observed covariates. (See Theorem 7.2 of \cite{Tsiatis2006} for technical details.)
Intuitively, these two models utilize more information than $P(R_{ij}^Y=1|A_i,\bfB_i)$ and $E[Y_{ij}|R_{ij}=1, A_i, \bfB_i]$ to predict $R_{ij}^Y$ and $Y_{ij}$, thereby further improving precision.
However, when $N_i$ is large, as in CRTs, these two distributions can take up to $2^{N_i}$ unique functions, which are often too many to be accurately learned from observed data. Therefore, while we acknowledge the existence of a more efficient estimator than $\widehat{\Delta}$ and $\widehat{\Delta}^{\textup{ml}}$ asymptotically, such an estimator is often infeasible to obtain in finite samples and impractical, and, therefore, we do not further pursue such an estimator.

\subsection{Modeling the propensity score for treatment assignment}
Modeling the propensity score for treatment assignment has been established as a technique to improve precision in randomized trials (for examples, \citealp{balzer2016adaptive, zeng2021propensity}).
While the distribution of $A_i$ given covariates is known to be $\pi$, constructing an estimator for it (and using the same data to estimate $\Delta$) will result in an extra term in the influence function of $\widehat\Delta$, and this extra term can be equivalently obtained by fitting an outcome regression model, which explains why this seemingly unnecessary model for treatment assignment can improve precision.

Motivated by this idea, we also consider modeling $P(A_i=1|\bfB_i) = \pi(\bfB_i;\btheta_A)$ via maximum likelihood estimation, and assume that $\pi(\bfB_i;\btheta_A) = \pi$ for some $\btheta_A$. For example, $\pi(\bfB_i;\btheta_A)$ can be a logistic regression model on intercept and cluster-summary variables. The estimator $\widehat{\mu}_a$ is then updated by substituting $\pi(\bfB_i;\widehat{\btheta}_A)$ for $\pi$, and the estimating functions $\bpsi(\bfO_i;\btheta)$ is updated to account for the estimation of $\btheta_A$. With this change, the following proposition characterizes the benefits of modeling treatment assignments.

\begin{proposition}\label{prop:4}
    Assume Assumptions~1-3, regularity conditions, and a correctly specified model $\kappa(A_i,\bfB_{i}; \btheta_R)$ for outcome missingness\footnote{In the sense that the estimating function for $\btheta_R$ has mean zero conditioning on $(A_i,\bfB_i)$, which automatically holds under generalized estimating equations or maximum likelihood estimation assuming conditional independence. 
    For arbitrary methods to estimate $\btheta_R$, this extra assumption rules out estimating functions that involve artificial terms, like $(A_i-\pi)\bfB_i$, beyond functions of $\btheta_R$ that is needed for parameter estimation.}. 
    Then, given any weight $\bw(N_i)$ and any outcome model $\eta(A_i,\bfB_{i}; \btheta_Y)$, modeling $\pi(\bfB_i;\btheta_A)$ does not increase the asymptotic variance of $\widehat{\Delta}$, and it does not change the asymptotic variance of $\widehat{\Delta}$ when the outcome model $\eta(A_i,\bfB_{i}; \btheta_Y)$ is correctly specified.
\end{proposition}

Proposition~\ref{prop:4} generalizes the results for inverse-probability weighted estimator from individually randomized trials to CRTs with missing data \citep{chang2023covariate, zhao2023covariate}. When the model for outcome missingness is correctly specified, adding a model for treatment assignment guarantees no efficiency loss asymptotically, which justifies the benefits of this technique in CRTs. For example, when $\eta(A_i,\bfB_{i}; \btheta_Y)$ is set to be zero, $\widehat{\Delta}$ reduces to the inverse-probability weighted estimator, and modeling $\pi(\bfB_i;\btheta_A)$ can improve precision as long as the covariates are prognostic. When there are no or few missing outcomes, modeling $\pi(\bfB_i;\btheta_A)$ can also supplement the outcome model to improve precision. 

\section{Proof}\label{asec: proof}
For notation convenience, we abbreviate the parametric models for nuisance functions \\ $\eta(a, \bfB_{i}; \btheta_Y)$ and $\kappa(a, \bfB_{i}; \btheta_Y)$ as $\eta_{ij}(a,\btheta_R)$ and $\kappa_{ij}(a,\btheta_Y)$, respectively. 
Furthermore, we denote $\eta_{ij}'(a, \btheta_Y^*)$ as the partial derivation of $\eta_{ij}(a, \btheta_Y)$ on $\btheta_Y$ evaluated at $\btheta_Y=\btheta_Y^*$; likewise, we denote $\kappa_{ij}'(a, \btheta_R^*)$ as the partial derivation of $\kappa_{ij}(a, \btheta_R)$ on $\btheta_R$  evaluated at $\btheta_R=\btheta_R^*$.
When the nuisance functions are estimated by machine learning methods, we abbreviate $\eta(a, \bfB_{i})$ and $\kappa(a, \bfB_{i})$ as $\eta_{ij}(a)$ and $\kappa_{ij}(a)$, respectively.

In addition,
we denote $w_j^*(N_i) = (\sum_{j=1}^{N_i} w_j(N_i))^{-1}w_j(N_i)$ as the scaled weight function, $\bw^*(N_i) = (w_1^*(N_i), \dots, w_{N_i}^*(N_i))^\top \in \mathbb{R}^{N_i}$, and $\bone_{N_i}$ as an $N_i$-dimensional column vector of ones.

Finally, while our interest lies in $\Delta$ as a function $(E[Y_{ij}|A_i=1], E[Y_{ij}|A_i=0])$, our proof centers on the asymptotic results for $(\widehat{\mu}_1, \widehat{\mu}_0)$. If we can prove their consistency and joint asymptotic normality, then the delta method will directly imply the consistency and asymptotic normality for $\widehat{\Delta}$, and the delta method will yield the sandwich variance estimator for $\widehat{\Delta}$. Therefore, we modify the notation of $\theta$ and $\psi$ to exclude $\Delta$ and $f(\mu_1,\mu_0) -\Delta$ from the parameters and estimating equations hereafter.  

\subsection{Proof of Theorem 1}
\begin{proof}[Proof of Theorem 1.]
    We write the estimating functions for parameters $\btheta = (\mu_1,\mu_0, \btheta_R, \btheta_Y)$ as 
    \begin{align*}
        \bpsi(\bfO_i; \btheta) = \left(\begin{array}{c}
          U(1, \bfO_i; \btheta_R,\btheta_Y)  -\mu_1   \\
          U(0, \bfO_i; \btheta_R,\btheta_Y)  -\mu_0   \\
          \bpsi_R(\bfO_i; \btheta_R)\\
          \bpsi_Y(\bfO_i; \btheta_Y)
        \end{array}\right),
    \end{align*}
    where, for $a=0,1$, we define
    \begin{align*}
    U(a,\bfO_i; \btheta_R,\btheta_Y)= \sum_{j=1}^{N_i} w_j^*(N_i) \left[ \frac{I\{A_i=a\}}{\pi^a(1-\pi)^{1-a}}     \frac{R_{ij}^Y \{Y_{ij} - \eta_{ij}(a, \btheta_Y)\}}{ \kappa_{ij}(a, \btheta_R)} + \eta_{ij}(a, \btheta_Y)\right],
    \end{align*}
     $\bpsi_R(\bfO_i; \btheta_R)$ is the estimating function for $\btheta_R$, and $\bpsi_Y(\bfO_i; \btheta_Y)$ is the estimating function for $\btheta_Y$. Therefore, $\widehat\btheta = (\widehat\mu_1,\widehat\mu_0, \widehat\btheta_R, \widehat\btheta_Y)$ is a solution to $\sum_{i=1}^m \bpsi(\bfO_i; \btheta) = \bzero$.

     By the regularity conditions, Example 19.8 of \cite{vaart_1998} implies that $\bpsi(\bfO_i; \btheta)$ is Glivenko-Cantelli, and, hence, by Theorem 5.9 of \cite{vaart_1998}, $\widehat\btheta \xrightarrow{P} \btheta^*$, where $\btheta^* = (\mu_1^*, \mu_0^*, \btheta_R^*, \btheta_Y^*)$. To show consistency, i.e., $E[Y_{ij}|A_i=a]= \mu_a^*$, we have 
     \begin{align*}
         \mu_a^* &= E[U(a,\bfO_i; \btheta_R^*,\btheta_Y^*)] \\
         &= E[E[U(a,\bfO_i; \btheta_R^*,\btheta_Y^*)|\bfB_i]] \\
         &= E\left[\sum_{j=1}^{N_i} w_j^*(N_i) E\left[ \frac{I\{A_i=a\}}{\pi^a(1-\pi)^{1-a}}     \frac{R_{ij}^Y \{Y_{ij} -\eta_{ij}(a, \btheta_Y^*)\}}{ \kappa_{ij}(a, \btheta_R^*)} + \eta_{ij}(a, \btheta_Y^*)\bigg|\bfB_i\right]\right] \\
         &= E\left[\sum_{j=1}^{N_i} w_j^*(N_i) E\left[ \frac{I\{A_i=a\}}{\pi^a(1-\pi)^{1-a}}     \frac{E[R_{ij}^Y \{Y_{ij} -\eta_{ij}(a, \btheta_Y^*)\}|A_i=a,\bfB_i]}{ \kappa_{ij}(a, \btheta_R^*)} + \eta_{ij}(a, \btheta_Y^*)\bigg|\bfB_i\right]\right] \\
        &= E\left[\sum_{j=1}^{N_i} w_j^*(N_i) E\left[   \frac{E[R_{ij}^Y \{Y_{ij} -\eta_{ij}(a, \btheta_Y^*)\}|A_i=a,\bfB_i]}{ \kappa_{ij}(a, \btheta_R^*)} + \eta_{ij}(a, \btheta_Y^*)\bigg|\bfB_i\right]\right],
     \end{align*}
     where the second last equation applies the tower law of conditional expectation given $A_i, \bfB_i$, and the last equation results from Assumption 2 stating $P(A_i=a|\bfB_i) =  \pi^a(1-\pi)^{1-a}$. Continuing the derivation, Assumption 3 implies that
     \begin{align*}
         E[R_{ij}^Y \{Y_{ij} -\eta_{ij}(a, \btheta_Y^*)\}|A_i=a,\bfB_i] = E[R_{ij}^Y|A_i=a,\bfB_i] \{E[Y_{ij}|A_i=a,\bfB_i] -\eta_{ij}(a, \btheta_Y^*)\}.
     \end{align*}
     If either $E[R_{ij}^Y|A_i=a,\bfB_{i}] = \kappa_{ij}(a, \btheta_R^*)$ or $E[Y_{ij}|A_i=a,\bfB_{i} ]= \eta_{ij}(a, \btheta_Y^*)$, direct algebra shows that
     \begin{align*}
         \mu_a^* = E\left[\sum_{j=1}^{N_i} w_j^*(N_i)E[Y_{ij}|A_i=a,\bfB_i]\right] = E\left[\sum_{j=1}^{N_i} w_j^*(N_i)E[E[Y_{ij}|A_i=a,\bfB_i]|A_i =a, N_i]\right].
     \end{align*}
     By Assumption 1 (2), $E[E[Y_{ij}|A_i=a,\bfB_i]|A_i=a, N_i]$ does not vary across $j$. Since $\sum_{j=1}^{N_i} w_j^*(N_i) = 1$ by definition, we obtain $\mu_a^* = E[E[Y_{ij}|A_i=a,N_i]]$. Using the fact that $A_i$ is independent of $N_i$, we obtain $\mu_a^* = E[Y_{ij}|A_i=a]$.      
     

     We next prove the asymptotic normality. By the regularity conditions, Theorem 5.41 of \cite{vaart_1998} implies that
\begin{equation}\label{eq: asymptotic-linearity}
    \sqrt{m}(\widehat{\btheta}-\btheta^*) = \frac{1}{\sqrt{m}}\sum_{i=1}^mE\left[\frac{\partial }{\partial \btheta} \bpsi(\bfO_i; \btheta) \big |_{\btheta = \btheta^*}\right]^{-1} \bpsi(\bfO_i;\btheta^*) + o_p(\bone).
\end{equation}
By the central limit theorem, $\sqrt{m}(\widehat{\btheta}-\btheta^*)  \xrightarrow{d} N(0, \mathbf{V}_{\btheta})$, where
\begin{align*}
    \mathbf{V}_{\btheta} = E\left[\frac{\partial }{\partial \btheta} \bpsi(\bfO_i; \btheta) \big |_{\btheta = \btheta^*}\right]^{-1} E[\bpsi(\bfO_i;\btheta^*)\bpsi(\bfO_i;\btheta^*)^\top]E\left[\frac{\partial }{\partial \btheta} \bpsi(\bfO_i; \btheta) \big |_{\btheta = \btheta^*}\right]^{-1\ T}.
\end{align*}
Following Lemma 3 in the Supplementary Material of \cite{wang2023model}, we obtain consistency of the sandwich variance estimator, i.e., $\widehat{\mathbf{V}}_{\btheta} \xrightarrow{P} \mathbf{V}_{\btheta}$. Since $\Delta  = f(\mu_1^*,\mu_0^*)$, by the continuous mapping theorem, $\sqrt{m}(\widehat{\Delta}-\Delta)  \xrightarrow{d} N(0, V_{\Delta})$ and $\widehat{Var}(\widehat\Delta) \xrightarrow{P} V_{\Delta}$. Finally, Slusky's Theorem implies the desired convergence result, which completes the proof.
\end{proof}

\subsection{Proof of Proposition 1 of the main paper}
    \begin{proof}[Proof of Proposition 1 of the main paper.]
        We inherit all notations from the proof of Theorem 1. By Equation~(\ref{eq: asymptotic-linearity}), direct algebra shows that
        \begin{align*}
              &\sqrt{m}(\widehat{\mu}_a - \mu_a^*)
              = \frac{1}{\sqrt{m}} \sum_{i=1}^m U(a,\bfO_i; \btheta_R^*,\btheta_Y^*) - \boldsymbol{q}_{1a}^\top \bpsi_R(\bfO_i;\btheta_R^*) - \boldsymbol{q}_{2a}^\top \bpsi_Y(\bfO_i;\btheta_Y^*) - \mu_a^* + o_p(1),
        \end{align*}
        where
        \begin{align*}
            \boldsymbol{q}_{1a}^\top &= E\left[\frac{\partial}{\partial \btheta_R} U(a,\bfO_i; \btheta_R,\btheta_Y^*)\Big|_{\btheta_R=\btheta_R^*}\right]^\top E\left[\frac{\partial}{\partial \btheta_R}\bpsi_R(\bfO_i;\btheta_R)\Big|_{\btheta_R=\btheta_R^*}\right]^{-1}, \\
            \boldsymbol{q}_{2a}^\top &= E\left[\frac{\partial}{\partial \btheta_Y} U(a,\bfO_i; \btheta_R^*,\btheta_Y)\Big|_{\btheta_Y=\btheta_Y^*}\right]^\top E\left[\frac{\partial}{\partial \btheta_Y}\bpsi_Y(\bfO_i;\btheta_Y)\Big|_{\btheta_Y=\btheta_Y^*}\right]^{-1}.
        \end{align*}
    Therefore, by defining
    \begin{align*}
        IF_{ij}(a,\bfO_i) &= \frac{I\{A_i=a\}}{\pi^a(1-\pi)^{1-a}}     \frac{R_{ij}^Y \{Y_{ij} - \eta_{ij}(a, \btheta_Y^*)\}}{ \kappa_{ij}(a, \btheta_R^*)} + \eta_{ij}(a, \btheta_Y^*)- \boldsymbol{q}_{1a}^\top \bpsi_R(\bfO_i;\btheta_R^*) - \boldsymbol{q}_{2a}^\top \bpsi_Y(\bfO_i;\btheta_Y^*)
    \end{align*}
    and $IF_i(a,\bfO_i) = (IF_{i1}(a,\bfO_i), \dots, IF_{iN_i}(a,\bfO_i))^\top$, we have
     \begin{equation*}
        \sqrt{m}(\widehat{\mu}_a - E[Y_{ij}|A_i=a])  = \frac{1}{\sqrt{m}} \sum_{i=1}^m \bw^*(N_i)^\top IF_{i}(a,\bfO_i) + o_p(1).
    \end{equation*}
    To obtain the optimal weights $\bw(N_i)^{\textup{opt}}$, we observe that $\widehat{\Delta}$ has asymptotic variance
    \begin{align*}
        v(\bw) =E\left[\left\{\sum_{a=0}^1 f_a'\bw^*(N_i)^\top IF_{i}(a,\bfO_i)\right\}\left\{\sum_{a=0}^1 f_a'\bw^*(N_i)^\top IF_{i}(a,\bfO_i)\right\}^\top\right]= E\left[\bw^*(N_i)^\top\mathbf{H}_i \bw^*(N_i)\right],
    \end{align*}
    where $\mathbf{H}_i = E\left[\left\{\sum_{a=0}^1 f_a'IF_{i}(a,\bfO_i)\right\}\left\{\sum_{a=0}^1 f_a' IF_{i}(a,\bfO_i)\right\}^\top\big|N_i\right]$. Letting  \\$\bw^*(N_i)^{\textup{opt}} = (\bone_{N_i}^\top\mathbf{H}_i^{-1} \bone_{N_i})^{-1}\mathbf{H}_i^{-1} \bone_{N_i}$, we have, for any $\bw^*(N_i)$ that sums up to 1,
    \begin{align*}
        E\left[\bw^*(N_i)^\top\mathbf{H}_i \bw^*(N_i)^{\textup{opt}}\right] = E\left[\frac{1}{\bone_{N_i}^\top\mathbf{H}_i^{-1} \bone_{N_i}}\right] = v(\bw^{\textup{opt}}),
    \end{align*}
    and, therefore,
    \begin{align*}
        0 &\le E\left[\left\{\bw^*(N_i) -\bw^*(N_i)^{\textup{opt}}\right\}^\top\mathbf{H}_i \left\{\bw^*(N_i) -\bw^*(N_i)^{\textup{opt}}\right\}\right] \\
        &= v(\bw) - 2v(\bw^{\textup{opt}}) + v(\bw^{\textup{opt}}) \\
        &= v(\bw) - v(\bw^{\textup{opt}}),
    \end{align*}
    which completes the proof that $\bw(N_i)^{\textup{opt}}$ leads to the smallest asymptotic variance for $\widehat{\Delta}$.
    \end{proof}

\subsection{Proof of Proposition A}
\begin{proof}[Proof of Proposition A.]
    We inherit all notations from the Proof of Theorem 1. When $P(R_{ij}^Y=1|A_i=a, \bfB_{i})$ is known, the estimating function for $\btheta_R$ disappears. As a special case, we follow the proof of Theorem 1 to obtain that
    \begin{align*}
        E\left[\frac{\partial }{\partial \btheta} \bpsi(\bfO_i; \btheta) \big |_{\btheta = \btheta^*}\right] &= \left(\begin{array}{ccc}
        -1   &  0 & E\left[\frac{\partial }{\partial \btheta_Y} U(1, \bfO_i; \btheta_Y) \big |_{\btheta_Y = \btheta_Y^*}\right]\\
        0    &  -1 & E\left[\frac{\partial }{\partial \btheta_Y} U(0, \bfO_i; \btheta_Y) \big |_{\btheta_Y = \btheta_Y^*}\right]\\
        0    &  0  & E\left[\frac{\partial }{\partial \btheta_Y} \bpsi_Y(\bfO_i; \btheta_Y) \big |_{\btheta_Y = \btheta_Y^*}\right]
        \end{array}\right)
    \end{align*}
    where
    \begin{align*}
        U(a,\bfO_i,\btheta_Y)= \sum_{j=1}^{N_i} w_j^*(N_i)  \left[ \frac{I\{A_i=a\}}{\pi^a(1-\pi)^{1-a}}     \frac{R_{ij}^Y \{Y_{ij} - \eta_{ij}(a, \btheta_Y)\}}{ P(R_{ij}^Y=1|A_i=a, \bfB_{i})} + \eta_{ij}(a, \btheta_Y)\right].
    \end{align*}
    Since $E[R_{ij}^Y|A_i,\bfB_{i}] = P(R_{ij}^Y=1|A_i, \bfB_{i})$ and $A_i$ is independent of $\bfB_i$, we have
    \begin{align*}
        & E\left[\frac{\partial }{\partial \btheta_Y} U(a, \bfO_i; \btheta_Y) \big |_{\btheta_Y = \btheta_Y^*}\right] \\
        &=-E\left[\sum_{j=1}^{N_i} w_j^*(N_i)\left\{ \frac{I\{A_i=a\}}{\pi^a(1-\pi)^{1-a}}  \frac{R_{ij}^Y }{ P(R_{ij}^Y=1|A_i=a, \bfB_{i})}-1\right\}\eta'_{ij}(a, \btheta_Y^*)\right]\\
        &=\bzero,
    \end{align*}
    by the tower law of conditional expectation. 
    Therefore, $E\left[\frac{\partial }{\partial \btheta} \bpsi(\bfO_i; \btheta) \big |_{\btheta = \btheta^*}\right]$ is a block-diagonal matrix, which implies that
    \begin{equation*}
        \sqrt{m}\left(\begin{array}{c}
            \widehat{\mu}_1 - \mu_1^* \\
            \widehat{\mu}_0 - \mu_0^* 
        \end{array}\right) = \frac{1}{\sqrt{m}}\sum_{i=1}^m \left(\begin{array}{c}
            U(1,\bfO_i,\btheta_Y^*) -  \mu_1^* \\
            U(0,\bfO_i,\btheta_Y^*) -  \mu_0^*
        \end{array} \right) + o_p(\bone),
    \end{equation*}
    and the asymptotic variance of $(\widehat{\mu}_1, \widehat{\mu}_0)$ is $Var(\bU(\bfO_i,\btheta_Y^*))$ for $\bU(\bfO_i,\btheta_Y^*) = (U(1,\bfO_i,\btheta_Y^*), U(0,\bfO_i,\btheta_Y^*))^\top$.
    
    We further denote
    \begin{align*}
        U^*(a,\bfO_i)= \sum_{j=1}^{N_i} w_j^*(N_i)  \left[ \frac{I\{A_i=a\}}{\pi^a(1-\pi)^{1-a}}     \frac{R_{ij}^Y \{Y_{ij} - E[Y_{ij}|A_i=a,\bfB_{i}]\}}{ P(R_{ij}^Y=1|A_i=a, \bfB_{i})} + E[Y_{ij}|A_i=a,\bfB_{i}]\right],
    \end{align*}
    and $\bU^*(\bfO_i) = (U^*(1,\bfO_i), U^*(0,\bfO_i))^\top$. Of note, $\bU^*(\bfO_i) = \bU(\bfO_i,\btheta_Y^*)$ if the outcome regression model is correctly specified. We can compute 
    \begin{align*}
        Var(\bU(\bfO_i,\btheta_Y^*)) &= Var(\bU^*(\bfO_i)) + 2Cov(\bU^*(\bfO_i), \bU(\bfO_i,\btheta_Y^*) - \bU^*(\bfO_i))\\
        &\quad + Var(\bU(\bfO_i,\btheta_Y^*) - \bU^*(\bfO_i)). \numberthis\label{eq: U-comparision}
    \end{align*}
    To draw the comparison between $Var(\bU(\bfO_i,\btheta_Y^*))$ and $Var(\bU^*(\bfO_i))$ in Equation~\eqref{eq: U-comparision}, we next calculate $Cov(\bU^*(\bfO_i), \bU(\bfO_i,\btheta_Y^*) - \bU^*(\bfO_i))$. To this end, we first observe that
    \begin{align*}
        & U(a,\bfO_i,\btheta_Y^*) - U^*(a,\bfO_i) \\
        &= \sum_{j=1}^{N_i} w_j^*(N_i) \left\{\frac{I\{A_i=a\}}{\pi^a(1-\pi)^{1-a}}  \frac{R_{ij}^Y }{ P(R_{ij}^Y=1|A_i=a, \bfB_{i})}-1\right\} \{E[Y_{ij}|A_i=a,\bfB_{i}] - \eta_{ij}(a, \btheta_Y^*)\} 
    \end{align*}
    is a function of $(\bR_i^Y, A_i, \bfB_i)$, and we further have
    \begin{align*}
        E[U(a,\bfO_i,\btheta_Y^*) - U^*(a,\bfO_i)|\bfB_i]
       &= E[E[U(a,\bfO_i,\btheta_Y^*) - U^*(a,\bfO_i)|A_i,\bfB_i]|\bfB_i] \\
       &= \sum_{j=1}^{N_i} w_j^*(N_i) \left\{\frac{P(A_i=a|\bfB_i)}{\pi^a(1-\pi)^{1-a}} -1\right\} \{E[Y_{ij}|A_i=a,\bfB_{i}] - \eta_{ij}(a, \btheta_Y^*)\} \\
       &= 0, \\
        E[U^*(a,\bfO_i)|\bR_i^Y, A_i, \bfB_i] &=\sum_{j=1}^{N_i} w_j^*(N_i) E[Y_{ij}|A_i=a,\bfB_{i}].
    \end{align*}
    Taken together, we get, for any $a,a' \in \{0,1\}$,
    \begin{align*}
        & Cov\{U^*(a,\bfO_i), U(a',\bfO_i,\btheta_Y^*) - U^*(a',\bfO_i)\}\\
        &=E[U^*(a,\bfO_i)\{U(a',\bfO_i,\btheta_Y^*) - U^*(a',\bfO_i)\}]\\
        &=E[E[U^*(a,\bfO_i)|\bR_i^Y, A_i, \bfB_i]\{U(a',\bfO_i,\btheta_Y^*) - U^*(a',\bfO_i)\}]\\        &=E\left[\sum_{j=1}^{N_i} w_j^*(N_i) E[Y_{ij}|A_i=a,\bfB_{i}] \{U(a',\bfO_i,\btheta_Y^*) - U^*(a',\bfO_i)\}\right] \\
        &=E\left[\sum_{j=1}^{N_i} w_j^*(N_i) E[Y_{ij}|A_i=a,\bfB_{i}] E[U(a',\bfO_i,\btheta_Y^*) - U^*(a',\bfO_i)|\bfB_i]\right] \\    
        &=0,\end{align*}
    which implies $Cov(\bU^*(\bfO_i), \bU(\bfO_i,\btheta_Y^*) - \bU^*(\bfO_i)) = \bzero$. Therefore, $Var(\bU(\bfO_i,\btheta_Y^*)) - Var(\bU^*(\bfO_i))$ is positive  semi-definite, and $Var(\bU(\bfO_i,\btheta_Y^*)) = Var(\bU^*(\bfO_i))$ if $E[Y_{ij}|A_i=a,\bfB_{i}] = \eta_{ij}(a, \btheta_Y^*)$. By the delta method, the asymptotic variance of $\widehat{\Delta}$ is minimized at $E[Y_{ij}|A_i=a,\bfB_{i}] = \eta_{ij}(a, \btheta_Y^*)$.

    When $P(R_{ij}^Y=1|A_i=a, \bfB_{i})$ is not known, but consistently estimated using independent data, the above derivation remains the same except for an extra step, where we first let the size of independent data goes to infinity. This proof is similar to the proof of Theorem 2 with cross-fitting, where a detailed proof is provided. 

    Finally, we prove the result assuming $(R_{i1},\dots, R_{iN_i})$ are mutually independent given $A_i,\bfB_i$, and  $P(R_{ij}^Y=1|A_i,\bfB_{i})=\kappa_{ij}(A, \btheta_R^*)$. Following a similar proof to the case with known $P(R_{ij}^Y=1|A_i, \bfB_{i})$, we have 
    \begin{equation*}
        \sqrt{m}\left(\begin{array}{c}
            \widehat{\mu}_1 - \mu_1^* \\
            \widehat{\mu}_0 - \mu_0^* 
        \end{array}\right) = \frac{1}{\sqrt{m}}\sum_{i=1}^m \left(\begin{array}{c}
            U(1,\bfO_i;\btheta_R^*,\btheta_Y^*) -  \mu_1^* - \boldsymbol{l}_{1}^\top \mathbf{D}^{-1}\bpsi_R(\bfO_i;\btheta_R^*) \\
            U(0,\bfO_i;\btheta_R^*,\btheta_Y^*) -  \mu_0^* - \boldsymbol{l}_{0}^\top \mathbf{D}^{-1}\bpsi_R(\bfO_i;\btheta_R^*)
        \end{array} \right) + o_p(\bone),
    \end{equation*}
    where 
    $\boldsymbol{l}_{a} = -E\left[\frac{\partial}{\partial \btheta_R} U(a,\bfO_i; \btheta_R,\btheta_Y^*)\Big|_{\btheta_R=\btheta_R^*}\right]$ and $\mathbf{D}= -E\left[\frac{\partial}{\partial \btheta_R}\bpsi_R(\bfO_i;\btheta_R)\Big|_{\btheta_R=\btheta_R^*}\right]$ with
        \begin{align*}
        U(a,\bfO_i;\btheta_R,\btheta_Y)= \sum_{j=1}^{N_i} w_j^*(N_i)  \left[ \frac{I\{A_i=a\}}{\pi^a(1-\pi)^{1-a}}     \frac{R_{ij}^Y \{Y_{ij} - \eta_{ij}(a, \btheta_Y)\}}{\kappa_{ij}(a, \btheta_R)} + \eta_{ij}(a, \btheta_Y)\right].
    \end{align*}
    By the assumption that $(R_{i1},\dots, R_{iN_i})$ are mutually independent given $(A_i,\bfB_i)$, the log-likelihood function for estimating $\btheta_R$ is $\sum_{j=1}^{N_i} R_{ij} \log(\kappa_{ij}(A_i, \btheta_R)) + (1-R_{ij})\log(1-\kappa_{ij}(A_i, \btheta_R))$, which implies 
    \begin{align*}
        \bpsi_R(\bfO_i;\btheta_R) = \sum_{j=1}^{N_i} \frac{R_{ij} - \kappa_{ij}(A_i, \btheta_R)}{\kappa_{ij}(A_i, \btheta_R)\{1-\kappa_{ij}(A_i, \btheta_R)\}}  \kappa_{ij}'(A_i, \btheta_R).
    \end{align*}
    Direct algebra implies that
    \begin{align*}
        \mathbf{D} &= E\left[\sum_{j=1}^{N_i} \frac{\kappa_{ij}'(A,\btheta_R^*)\kappa_{ij}'(A,\btheta_R^*)^\top}{\kappa_{ij}(A,\btheta_R^*)(1-\kappa_{ij}(A,\btheta_R^*))}\right] \\
        \boldsymbol{l}_{a} &= E\left[\sum_{j=1}^{N_i} w_{j}^*(N_i)\frac{\kappa_{ij}'(a,\btheta_R^*)}{\kappa_{ij}(a,\btheta_R^*)} \left\{E[Y_{ij}|A_i=a,\bfB_i] - \eta_{ij}(a,\btheta_Y^*)\right\}\right].
    \end{align*}
    Since we further assumed that $R_{ij}$ is independent of $R_{ij'}$ given $A_i, \bfB_i$, we have
    \begin{align*}
    &E[\bpsi_R(\bfO_i;\btheta_R^*)\bpsi_R(\bfO_i;\btheta_R^*)^\top]\\
    &= E\left[\sum_{j=1}^{N_i} \frac{(R_{ij}- \kappa_{ij}(A,\btheta_R^*))^2\kappa_{ij}'(A,\btheta_R^*)\kappa_{ij}'(A,\btheta_R^*)^\top}{\kappa_{ij}^2(A,\btheta_R^*)(1-\kappa_{ij}(A,\btheta_R^*))^2}\right] \\
        &= E\left[\sum_{j=1}^{N_i} \frac{\kappa_{ij}'(A,\btheta_R^*)\kappa_{ij}'(A,\btheta_R^*)^\top}{\kappa_{ij}(A,\btheta_R^*)(1-\kappa_{ij}(A,\btheta_R^*))}\right]\\
        &= \mathbf{D},\numberthis\label{def:math_D}\\
    &E[\bpsi_R(\bfO_i;\btheta_R^*)\{ U(a,\bfO_i,\btheta_Y^*) -  \mu_a^*\}] \\
    &= E\left[\frac{I\{A_i=a\}}{\pi^a(1-\pi)^{1-a}}\sum_{j=1}^{N_i}w_{ij}^*\frac{R_{ij}(R_{ij} - \kappa_{ij}(A,\btheta_R^*))\kappa_{ij}'(A,\btheta_R^*)}{\kappa_{ij}^2(A,\btheta_R^*)(1-\kappa_{ij}(A,\btheta_R^*))}\{Y_{ij}  -\eta_{ij}(a,\btheta_Y^*)\}\right] \\
    &=E\left[\sum_{j=1}^{N_i} w_{ij}^*\frac{\kappa_{ij}'(a,\btheta_R^*)}{\kappa_{ij}(a,\btheta_R^*)} \left\{E[Y_{ij}|A_i=a,\bfB_i]- \eta_{ij}(a,\btheta_Y^*)\right\}\right]\\
    &=\boldsymbol{l}_{a}.\numberthis\label{def:l_a}
    \end{align*}
    Denoting $\bU(\bfO_i;\btheta_Y^*) = ( U(1,\bfO_i;\btheta_Y^*),  U(0,\bfO_i;\btheta_Y^*))^\top$ and $\boldsymbol{l} = (\boldsymbol{l}_{1}, \boldsymbol{l}_{0})$, we use the above facts to obtain that the asymptotic variance of $(\widehat{\mu}_1, \widehat{\mu}_0)$ is
    \begin{align*}
        \bfV_{\mu} &= Var\{\bU(\bfO_i;\btheta_Y^*) - \boldsymbol{l}^\top \mathbf{D}^{-1}\bpsi_R(\bfO_i;\btheta_R^*)\} \numberthis\label{def: v_mu_prop2}\\
        &= Var\{\bU(\bfO_i;\btheta_Y^*)\} - \boldsymbol{l}^\top \mathbf{D}^{-1} \boldsymbol{l}.
    \end{align*}
    This result implies that, for any outcome model, using a correctly specified model to learn the missing propensity score does not harm the precision. 

    Given the formula of $\bfV_{\mu}$, the final step is to show $\bfV_{\boldsymbol{\mu}} - Var(\bU^*(\bfO_i))$ is positive semi-definite, thereby proving that a correctly specified outcome model will minimize the variance. 
    To this end, we recall  Equation~(\ref{eq: U-comparision}) showing $Var(\bU(\bfO_i;\btheta_Y^*)) = Var(\bU^*(\bfO_i)) + Var\{\bU(\bfO_i;\btheta_Y^*)-\bU^*(\bfO_i)\}$. 
    Therefore, the definition of  $\mathbf{D}$ and $\boldsymbol{l}$ (Equations~\eqref{def:math_D} and \eqref{def:l_a}) implies that
        \begin{align*}
        \bfV_{\boldsymbol{\mu}} &= Var(\bU(\bfO_i;\btheta_Y^*)) - \boldsymbol{l}^\top \mathbf{D}^{-1} \boldsymbol{l} \\
        &= Var(\bU^*(\bfO_i)) + Var\{\bU(\bfO_i;\btheta_Y^*)-\bU^*(\bfO_i)\}  - \boldsymbol{l}^\top \mathbf{D}^{-1} \boldsymbol{l} \\
        &= Var(\bU^*(\bfO_i)) + Var\{\bU(\bfO_i;\btheta_Y^*)-\bU^*(\bfO_i) - \boldsymbol{l}^\top \mathbf{D}^{-1} \bpsi_R(\bfO_i;\btheta_R^*) \} \\
        &\quad + 2\left[Cov\left\{\bU(\bfO_i;\btheta_Y^*)-\bU^*(\bfO_i), \boldsymbol{l}^\top \mathbf{D}^{-1} \bpsi_R(\bfO_i;\btheta_R^*)\right\} - \boldsymbol{l}^\top \mathbf{D}^{-1}\boldsymbol{l}\right] \\
        &=Var(\bU^*(\bfO_i)) + Var\{\bU(\bfO_i;\btheta_Y^*)-\bU^*(\bfO_i) - \boldsymbol{l}^\top \mathbf{D}^{-1} \bpsi_R(\bfO_i;\btheta_R^*) \} \\
            &\quad -2Cov\left\{\bU^*(\bfO_i), \boldsymbol{l}^\top \mathbf{D}^{-1} \bpsi_R(\bfO_i;\btheta_R^*)\right\}.
    \end{align*}
    If we can show that $Cov\left\{\bU^*(\bfO_i), \bpsi_R(\bfO_i;\btheta_R^*)\right\} = \bzero$, then we obtain the desired result that $\bfV_{\boldsymbol{\mu}} - Var(\bU^*(\bfO_i))$ is positive semi-definite. To show this, we have
    \begin{align*}
    &E[\bpsi_R(\bfO_i;\btheta_R^*)\{ U^*(a,\bfO_i) -  \mu_a^*\}] \\
 &=E[\bpsi_R(\bfO_i;\btheta_R^*)E[ U^*(a,\bfO_i) -  \mu_a^*|\bfR_i, A_i, \bfB_i]] \\
 &= E\left[\bpsi_R(\bfO_i;\btheta_R^*)\sum_{j=1}^{N_i} w_j^*(N_i) (E[Y_{ij}|A_i=a,\bfB_{i}]-\mu_a^*)\right] \\
 &= E\left[E[\bpsi_R(\bfO_i;\btheta_R^*)|A_i,\bfB_i]\sum_{j=1}^{N_i} w_j^*(N_i) (E[Y_{ij}|A_i=a,\bfB_{i}]-\mu_a^*)\right]\\
 &= E\left[\bzero \times \sum_{j=1}^{N_i} w_j^*(N_i) (E[Y_{ij}|A_i=a,\bfB_{i}]-\mu_a^*)\right] \\
 &= \bzero,
    \end{align*}
where the second last equation results from the fact that the outcome missing model $\kappa$ is correctly specified. 
    When the outcome model is correctly specified, i.e., $E[Y_{ij}|A_i=a,\bfB_i] = \eta_{ij}(A,\btheta_Y^*)$, we have $\bU^*(\bfO_i)-\bU(\bfO_i;\btheta_Y^*) = \bzero$ and, therefore, $\bfV_{\boldsymbol{\mu}} = Var(\bU^*(\bfO_i))$, implying that the smallest variance is attained when the outcome model is correctly specified.
    
\end{proof}
\subsection{Proof of Theorem 2}
Denote $\mathbb{P}_m X = m^{-1} \sum_{i=1}^m X_i$ and $\mathbb{G}_m = m^{1/2}\{\mathbb{P}_m X - E[X]\}$ for i.i.d. samples from a distribution $X$. Furthermore, we define $\mathbb{P}_k X = |\mathcal{Q}_k|^{-1} \sum_{i\in \mathcal{Q}_k} X_i$ and $\mathbb{G}_k 
 = |\mathcal{Q}_k|^{1/2} (\mathbb{P}_k X - E[X])$, where $|\mathcal{Q}_k|$ is the size of fold $k$. Without loss of generality, we assume that all $K$ folds have the same size. In addition, we denote
 \begin{align*}
     U_a(\kappa, \eta) =  \sum_{j=1}^N w_j^*(N_i) \frac{I\{A=a\}}{\pi^{a}(1-\pi)^{1-a}} \frac{R_{.j}\{Y_{.j} - \eta_{.j}(a)\}}{\kappa_{.j}(a)} + \eta_{.j}(a).
 \end{align*}
 Given the above definitions, we have $\widehat{\mu}_a^{\textup{ml}} = K^{-1} \sum_{k=1}^K \mathbb{P}_k U_a(\widehat{\kappa}^{(k)}, \widehat{\eta}^{(k)})$. Denoting the probability limit of $(\widehat{\kappa}^{(k)}, \widehat{\eta}^{(k)})$ as $(\kappa^*, \eta^*)$, Assumption 5 implies that $\kappa^* = P(R_{ij}=1|A_i=a, \bfB_i)$ and $\eta^* = E[Y_{ij}|R_{ij}=1,A_i=a, \bfB_i]$. Then $E[Y_{ij}|A_i=a]  = E[U_a(\kappa^*, \eta^*)]$ following the proof of Theorem 1. Next, we prove the asymptotic normality. Direct algebra shows that 
\begin{align*}
    & m^{1/2}(\widehat{\mu}_a^{\textup{ml}}  - E[Y_{ij}|A_i=a]) \\
    &=m^{1/2}K^{-1} \sum_{k=1}^K \mathbb{P}_k U_a(\widehat{\kappa}^{(k)}, \widehat{\eta}^{(k)}) - m^{1/2}E[U_a(\kappa^*, \eta^*)] \\
    &= K^{-1/2}\sum_{k=1}^K\left[\mathbb{G}_kU_a(\kappa^*, \eta^*) + \underbrace{\mathbb{G}_k\{U_a(\widehat{\kappa}^{(k)}, \widehat{\eta}^{(k)}) -U_a(\kappa^*, \eta^*)\}}_{\textup{denoted as}\,\, J_1}\right.\\ 
    &\quad + \underbrace{(\frac{m}{K})^{1/2} E[U_a(\widehat{\kappa}^{(k)}, \widehat{\eta}^{(k)}) -U_a(\kappa^*, \eta^*)]}_{\textup{denoted as}\,\, J_2}\bigg].
\end{align*}
In the above expression, we denote $J_1 = \mathbb{G}_k\{U_a(\widehat{\kappa}^{(k)}, \widehat{\eta}^{(k)}) -U_a(\kappa^*, \eta^*)\}$ and\\ $J_2 = (\frac{m}{K})^{1/2} E[U_a(\widehat{\kappa}^{(k)}, \widehat{\eta}^{(k)}) -U_a(\kappa^*, \eta^*)]$. If we can show that $J_1 = o_p(1)$ and $J_2 = o_p(1)$, then, since $K$ is fixed, we have
$$m^{1/2}(\widehat{\mu}_a^{\textup{ml}}  - E[Y_{ij}|A_i=a]) =  K^{-1/2}\sum_{k=1}^K \mathbb{G}_kU_a(\kappa^*, \eta^*) + o_p(1) = \mathbb{G}_m U_a(\kappa^*, \eta^*) + o_p(1)$$
for $a=0,1$. Then, by the delta method, the asymptotic variance of $\Delta^{\textup{ml}}$ will have the asymptotic variance $Var\{f_1'U_1(\kappa^*, \eta^*) + f_0'U_0(\kappa^*, \eta^*)\}$, which $v^{opt}$ defined in Equation (4) of the main paper.

We next show $J_1 = o_p(1)$. By Chebyshev's inequality, $J_1 = o_p(1)$ given $\mathcal{Q}_{-k}$ \\ if $E[\{U_a(\widehat{\kappa}^{(k)}, \widehat{\eta}^{(k)}) -U_a(\kappa^*, \eta^*)\}^2|\mathcal{Q}_{-k}]=o_p(1)$. Since we assume that $\widehat{\eta}^{(k)} - \eta^* = o_p(1)$ in $L_2$-norm, Markov's inequality implies that $E[(\widehat{\eta}^{(k)}-\eta^*)^2|\mathcal{Q}_{-k}] = o_p(1)$. since $\widehat{\eta}^{(k)}$ is fixed conditioning on $\mathcal{Q}_{-k}$, we have
\begin{align*}
     E[\{U_a(\widehat{\kappa}^{(k)}, \widehat{\eta}^{(k)}) -U_a(\kappa^*, \eta^*)\}^2|\mathcal{Q}_{-k}]&= E\left[\left\{\sum_{j=1}^{N}  w_{j}^*\frac{(AR_{.j}-\pi_a\widehat{\kappa}_j^{(k)})(\eta_j^*-\widehat{\eta}_j^{(k)})}{\pi_a\widehat{\kappa}_j^{(k)}}\right\}^2\bigg|\mathcal{Q}_{-k}\right] \\
    &\le E\left[\sum_{j=1}^{N} N(w_{j}^*)^2\frac{(AR_{.j}-\pi_a\widehat{\kappa}_j^{(k)})^2(\eta_j^*-\widehat{\eta}_j^{(k)})^2}{(\pi_a\widehat{\kappa}_j^{(k)})^2}\bigg|\mathcal{Q}_{-k}\right] \\
    &= O_p(1) E\left[\sum_{j=1}^N (w_{j}^*)^2 (\eta_j^*-\widehat{\eta}_j^{(k)})^2 \bigg|\mathcal{Q}_{-k}\right] \\
    &= O_p(1) o_p(1)\\
    &= o_p(1),
\end{align*}
 where $w_j^* = w_j^*(N)$ and $\pi_a = \pi^a(1-\pi)^{1-a}$. In the above derivation, the first inequality results from the Cauchy-Schwarz inequality, the next equality uses Assumption 5(2) that $\kappa_j^{(k)}$ is uniformly bounded and Assumption 5(3) that $N_i$ is upper bounded, and the second last equality results from $w_j^* \le 1$ and $E[(\widehat{\eta}^{(k)}-\eta^*)^2|\mathcal{Q}_{-k}] = o_p(1)$. Therefore, $J_1 = o_p(1)$ given $\mathcal{Q}_{-k}$, which implies $J_1 = o_p(1)$ marginally.

 For $J_2$ we have
 \begin{align*}
     & (\frac{m}{K})^{1/2} E[U_a(\widehat{\kappa}^{(k)}, \widehat{\eta}^{(k)}) -U_a(\kappa^*, \eta^*)|\mathcal{Q}_{-k}, N]\\
     &= (\frac{m}{K})^{1/2} \sum_{j=1}^{N}  w_{j}^*E\left[\frac{(AR_{.j}-\pi_a\widehat{\kappa}_j^{(k)})(\eta_j^*-\widehat{\eta}_j^{(k)})}{\pi_a\widehat{\kappa}_j^{(k)}}\bigg|\mathcal{Q}_{-k}, N\right] \\
     &= (\frac{m}{K})^{1/2} \sum_{j=1}^{N}  w_{j}^* E\left[\frac{(\kappa_j^*-\widehat{\kappa}_j^{(k)})(\eta_j^*-\widehat{\eta}_j^{(k)})}{\widehat{\kappa}_j^{(k)}}\bigg|\mathcal{Q}_{-k}, N\right] \\
     &\le (\frac{m}{K})^{1/2} \sum_{j=1}^{N}  w_{j}^* E\left[\frac{(\kappa_j^*-\widehat{\kappa}_j^{(k)})^2}{(\widehat{\kappa}_j^{(k)})^2} \bigg|\mathcal{Q}_{-k},N\right]E\left[(\eta_j^*-\widehat{\eta}_j^{(k)})^2\bigg|\mathcal{Q}_{-k},N\right] \\
     &=(\frac{m}{K})^{1/2} \sum_{j=1}^{N}  w_{j}^* O_p(1) o_p(m^{-1/4})o_p(m^{-1/4}) \\
     &= o_p(1),
 \end{align*}
where the inequality uses Cauchy-Schwarz inequality, the following equality uses Assumptions 5(1) and (2), and the last equality comes from Assumption 5(3) that $N$ is upper bounded by a constant and the fact that $K$ is a fixed constant. 
Therefore, we successfully proved $J_2 = o_p(1)$. As a result, we showed the consistency and asymptotic normality of $\widehat{\mu}_a^{\textup{ml}}$, and, consequently, $\widehat{\Delta}^{\textup{ml}}$ via delta method.

Finally, we construct the variance estimator and prove its consistency. Defining
\begin{align*}
    \bU(\kappa,\eta) &= (U_1(\kappa,\eta), U_0(\kappa,\eta))^\top, \\
    \widehat{\boldsymbol{\mu}}_k^{\textup{ml}} &= \mathbb{P}_k \bU(\widehat{\kappa}^{(k)},\widehat{\eta}^{(k)}), \\
    \widehat{\bfSigma}_k &= \mathbb{P}_k  \{\bU(\widehat{\kappa}^{(k)},\widehat{\eta}^{(k)})- \widehat{\boldsymbol{\mu}}_k^{\textup{ml}}\}  \{\bU(\widehat{\kappa}^{(k)},\widehat{\eta}^{(k)})- \widehat{\boldsymbol{\mu}}_k^{\textup{ml}}\}^\top,
\end{align*}
we construct the variance estimator for $\boldsymbol{\mu}^* =(E[Y_{ij}|A_i=1], E[Y_{ij}|A_i=0])$ as
\begin{align*}
    \widehat{\bfV}_{\boldsymbol{\mu}}^{\textup{ml}} = \frac{1}{K} \sum_{k=1}^K \widehat{\bfSigma}_k,
\end{align*}
and the $\widehat{Var}(\widehat{\Delta}^{\textup{ml}})$ is obtained by Delta method based $\widehat{\bfV}_{\boldsymbol{\mu}}^{\textup{ml}}$. To prove the consistency of the variance estimator, it suffices to prove $\widehat{\bfSigma}_k \xrightarrow{P} \bfSigma$, where $\bfSigma = Var\{\bU(\kappa^*, \eta^*)\}$. To this end, we define $\underline{\bfSigma}_k = \mathbb{P}_k  \{\bU(\kappa^*,\eta^*)- \boldsymbol{\mu}^*\}\{\bU(\kappa^*,\eta^*)- \boldsymbol{\mu}^*\}^\top$, and by the law of large numbers, $\underline{\bfSigma}_k = \bfSigma +o_p(1)$ since $\bfSigma$ is finite. Therefore, it suffices to show $\widehat{\bfSigma}_k - \underline{\bfSigma} = o_p(1)$. 
Defining $\boldsymbol{L}_k = \bU(\widehat{\kappa}^{(k)},\widehat{\eta}^{(k)})- \widehat{\boldsymbol{\mu}}_k^{\textup{ml}} -\bU(\kappa^*,\eta^*)+ \boldsymbol{\mu}^*$, we have
\begin{align*}
    \widehat{\bfSigma}_k - \underline{\bfSigma}_k = \mathbb{P}_k \{\boldsymbol{L}_k\boldsymbol{L}_k^\top  + \boldsymbol{L}_k(\bU(\kappa^*,\eta^*)- \boldsymbol{\mu}^*)^\top +(\bU(\kappa^*,\eta^*)- \boldsymbol{\mu}^*)\boldsymbol{L}_k^\top \}.
\end{align*}
By Holder's inequality and the matrix norm, 
\begin{align*}
    || \widehat{\bfSigma}_k - \underline{\bfSigma}_k||_2^2 &\le (\mathbb{P}_k \boldsymbol{L}_k^\top\boldsymbol{L}_k)^2 + 2(\mathbb{P}_k \boldsymbol{L}_k^\top\boldsymbol{L}_k) (\mathbb{P}_k(\bU(\kappa^*,\eta^*)- \boldsymbol{\mu}^*)^\top(\bU(\kappa^*,\eta^*)- \boldsymbol{\mu}^*)) \\
    &= (\mathbb{P}_k \boldsymbol{L}_k^\top\boldsymbol{L}_k)^2 + 2(\mathbb{P}_k \boldsymbol{L}_k^\top\boldsymbol{L}_k) tr(\underline{\bfSigma}_k).
\end{align*}
Since $\underline{\bfSigma}_k = \bfSigma + o_p(1)$, then $tr(\underline{\bfSigma}_k) = O_p(1)$, which means it suffices to show $\mathbb{P}_k \boldsymbol{L}_k^\top\boldsymbol{L}_k = o_p(1)$, which is true by the following derivation:
\begin{align*}
    \mathbb{P}_k \boldsymbol{L}_k^\top\boldsymbol{L}_k &= \mathbb{P}_k ||\bU(\widehat{\kappa}^{(k)},\widehat{\eta}^{(k)})- \widehat{\boldsymbol{\mu}}_k^{\textup{ml}} -\bU(\kappa^*,\eta^*)+ \boldsymbol{\mu}^*||_2^2 \\
    &\le 2 \mathbb{P}_k||\bU(\widehat{\kappa}^{(k)},\widehat{\eta}^{(k)}) -\bU(\kappa^*,\eta^*)||_2^2 + 2||\widehat{\boldsymbol{\mu}}_k^{\textup{ml}}-\boldsymbol{\mu}^*||_2^2 \\
    &=2 E[||\bU(\widehat{\kappa}^{(k)},\widehat{\eta}^{(k)}) -\bU(\kappa^*,\eta^*)||_2^2| \mathcal{Q}_{-k}] + o_p(1) + 2o_p(1) \\
    &=o_p(1),
\end{align*}
where the third line results from the law of large numbers on $||\bU(\widehat{\kappa}^{(k)},\widehat{\eta}^{(k)}) -\bU(\kappa^*,\eta^*)||_2^2$ and $\widehat{\boldsymbol{\mu}}_k^{\textup{ml}}$, and the last line was already proved during proving $J_1 = o_p(1)$. Therefore, we obtain the desired convergence result by applying Slusky's Theorem.

\subsection{Proof of Proposition B}
\begin{proof}
We first derive the identification formula for $E[Y_{ij}|A_i=a]$ under Assumptions 1-3.
We have
\begin{align*}
    E[Y_{ij}|A_i=a]
    &= E[E[Y_{ij}|A_i=a, N_i]] \\
    &= E\left[\frac{1}{N_i}\sum_{j=1}^{N_i} E[Y_{ij}|A_i=a, N_i]\right] \\
    &=E\left[\frac{1}{N_i}\sum_{j=1}^{N_i} E[E[Y_{ij}|A_i=a,\bfB_i]|N_i]\right] \\
    &=E\left[\frac{1}{N_i}\sum_{j=1}^{N_i} E[Y_{ij}|R_{ij}=1,A_i=a,\bfB_i]\right].
\end{align*}

We next derive the efficient influence function. Let $\mathcal{P}$ denote the distribution for the observed data $\bO$, $E[Y_{ij}|A_i=a] = \Psi(\mathcal{P})$ and a parametric submodel $\mathcal{P}_t = t \mathcal{P}^* + (1-t)\mathcal{P}$ for $t \in [0,1]$, where $\mathcal{P}^*$ is a point-mass at $o_i$ in the support of $\mathcal{P}$. Furthermore, let $f$ denote the density of $\mathcal{P}$, $1_{o^*}(o)$ denote the Dirac delta function for $o^*$, i.e., the density of $\mathcal{P}^*$, and $f_t = t 1_{o^*}(o) + (1-t) f$. In addition, we denote $\bfB_{i}^- = \bfB_{i}\setminus\{N_i\}$. Then
\begin{align*}
     &\frac{d \Psi(\mathcal{P}_t)}{d t}\bigg|_{t=0}\\
     &= \frac{d}{dt} \int_n \frac{1}{n} \sum_{j=1}^n\int_{y_{.j},\bb^-} y_{.j} f_t(y_{.j}|r_{.j}^Y = 1,a, \bb^-, n) dy_{.j} f_t(\bb^-| n) d\bb^- f_t(n) dn \bigg|_{t=0} \\
     &=\int_n \frac{1}{n} \sum_{j=1}^n \int_{y_{.j},\bb^-} y_{.j} f(y_{.j}|r_{.j}^Y = 1,a, \bb^-, n) dy_{.j} f(\bb^-| n) d\bb^- f(n) dn \\
     &\left\{\frac{1_{y_{.j}^*, r_{.j}^Y{}^*,a^*, \bb^-{}^*, n^*}(y_{.j},1,a,\bb^-, n)}{f(y_{.j},1,a,\bb^-, n)} - \frac{1_{ r_{.j}^Y{}^*,a^*, \bb^-{}^*, n^*}(1,a,\bb^-, n)}{f(1,a,\bb^-, n)} + \frac{1_{\bb^-{}^*, n^*}(\bb^-, n)}{f(\bb^-, n)} - \frac{1_{n^*}(n)}{f(n)} +  \frac{1_{n^*}(n)}{f(n)} - 1\right\} \\
     &= \frac{1}{n^*} \sum_{j=1}^n \left[\frac{1_{r_{.j}^Y{}^*,a^*}(1,a)}{P(R_{.j}^Y=1, A=a|\bfB^-=\bb^-{}^*, n = n^*)}\{y_{.j}^* - E[Y_{.j}|R_{.j}^Y=1, A=a, \bfB^-=\bb^-{}^*, n = n^*]\}\right. \\
     & + E[Y_{.j}|R_{.j}^Y=1, A=a, \bfB^-=\bb^-{}^*, n = n^*] - \Psi(\mathcal{P})\bigg].
\end{align*}
By Assumption 2, $P(R_{.j}^Y=1, A=a|\bfB^-=\bb^-{}^*, n = n^*) = \pi^{a}(1-\pi)^{1-a} P(R_{.j}^Y=1|A=a,\bfB^-=\bb^-{}^*, n = n^*)$. Since $1_{r_{.j}^Y{}^*,a^*}(1,a) = I\{r_{.j}^Y{}^* = 1, a^*=a\}$, we obtain the desired efficient influence function for $E[Y_{ij}|A_i=a]$ by substituting $(Y_{ij}, R_{ij}^Y, A_i, \bfB_{i})$ for $(y_{.j}^*, r_{.j}^Y{}^*,a^*, \bb^-{}^*, n^*)$, which is
\begin{align*}
    EIF(a)&= \frac{1}{N} \sum_{j=1}^N \frac{I\{A=a\}}{\pi^{a}(1-\pi)^{1-a}} \frac{R_{.j}(Y_{.j} - E[Y_{.j}|R_{.j}=1, A=a,\bfB])}{P(R_{.j}=1| A=a,\bfB)} + E[Y_{.j}|R_{.j}=1, A=a,\bfB] - \Psi(\mathcal{P}).
\end{align*}
We then verify that this parameter belongs to the nuisance tangent space defined by our assumptions. By Assumptions 1-3 and the additional assumption that $(R_{ij}, Y_{ij})$ are independent given $A_i,\bfB_i$, we have
\begin{align*}
    \mathcal{P}^{\bO} = \left\{\mathcal{P}^{R_{.j}Y_{.j}|A,\bfB}\mathcal{P}^{R_{.j}|A,\bfB}\right\}^N\mathcal{P}^{\bfB} \mathcal{P}^A ,
\end{align*}
which implies the nuisance tangent space is
\begin{align*}
    \mathcal{T}(\bO) &= \left\{\sum_{j=1}^N S_{RY}(R_{.j}Y_{.j},A,\bfB) + S_{R}(R_{.j},A,\bfB) + S_{\bfB}(\bfB): S_{RY}, S_R, S_{\bfB} \in L^2(\mathcal{P})\right.\\
    &\quad E[S_{RY}(R_{.j}Y_{.j},A,\bfB)|A,\bfB] = E[S_{R}(R_{.j},A,\bfB)|A,\bfB] = E[S_{\bfB}(\bfB)] = 0\bigg\}.
\end{align*}
To show $EIF(a) \in \mathcal{T}(\bO)$, we define 
\begin{align*}
    S_{RY}(R_{.j}Y_{.j},A,\bfB)  &=  \frac{I\{A=a\}}{N\pi^{a}(1-\pi)^{1-a}} \left\{\frac{R_{.j}Y_{.j}}{P(R_{.j}=1| A=a,\bfB)} - E[Y_{.j}|R_{.j}=1, A=a,\bfB]\right\},\\
    S_{R}(R_{.j},A,\bfB) &= \frac{I\{A=a\}}{N\pi^{a}(1-\pi)^{1-a}} \left\{1-\frac{R_{.j}}{P(R_{.j}=1| A=a,\bfB)} \right\}E[Y_{.j}|R_{.j}=1, A=a,\bfB],\\
    S_{\bfB}(\bfB) &= E[Y_{.j}|R_{.j}=1, A=a,\bfB] - \Psi(\mathcal{P}),
\end{align*}
and it is straightforward that $EIF(a) =  S_{RY}(R_{.j}Y_{.j},A,\bfB) 
 + S_{R}(R_{.j},A,\bfB) +  S_{\bfB}(\bfB)$ and $E[S_{RY}(R_{.j}Y_{.j},A,\bfB)|A,\bfB] = E[S_{R}(R_{.j},A,\bfB)|A,\bfB] = E[S_{\bfB}(\bfB)] = 0$, which completes the proof that $EIF(a)$ is indeed the efficient influence function for $E[Y_{ij}|A_i=a]$.

 By the chain rule, the efficient influence function for $\Delta$ is then $f'_1 EIF(1) + f'_0 EIF(0)$, which implies that the efficiency lower bound for $\Delta$ is $Var\{f'_1 EIF(1) + f'_0 EIF(0)\}$. This is $v^{opt}$ defined in Equation (4) with $w_j(N_i) = 1$. Of note, when $w_j(N_i)$ is different across $j$, then $EIF(a)$ is no longer the efficient influence function for $E[Y_{ij}|A_i=a]$ since we assume that the $(R_{ij}, Y_{ij})$ are conditionally identically distributed and the resulting formula will not reside in the nuisance tangent space $\mathcal{T}(\bO)$.

\end{proof}
\subsection{Proof of Proposition C}
\begin{proof}[Proof of Proposition C.]
We inherit all notations from the proof of Proposition A. 
By Equation~(\ref{eq: asymptotic-linearity}), direct algebra shows that
        \begin{align*}
              \sqrt{m}(\widehat{\mu}_a - \mu_a^*)
              &= \frac{1}{\sqrt{m}} \sum_{i=1}^m U(a,\bfO_i;\btheta_R^*,\btheta_Y^*, \btheta_A^*) - \mu_a^* - \boldsymbol{q}_{1a}^\top \bpsi_R(\bfO_i;\btheta_R^*)\\
              &\quad - \boldsymbol{q}_{2a}^\top \bpsi_Y(\bfO_i;\btheta_Y^*) - \boldsymbol{q}_{3a}^\top \bpsi_A(\bfB_i;\btheta_A^*)  + o_p(1),
        \end{align*}
        where $\bpsi_A(\bfO_i; \btheta_A)$ is the estimating functions for $\btheta_A$ and
        \begin{align*}
            U(a,\bfO_i;\btheta_R,\btheta_Y, \btheta_A)&= \sum_{j=1}^{N_i} w_j^*(N_i)  \left[ \frac{I\{A_i=a\}}{\pi(\bfB_i;\btheta_A)^a(1-\pi(\bfB_i;\btheta_A))^{1-a}}     \frac{R_{ij}^Y \{Y_{ij} - \eta_{ij}(a, \btheta_Y)\}}{\kappa_{ij}(a,\btheta_R)} + \eta_{ij}(a, \btheta_Y)\right],\\
            \boldsymbol{q}_{1a}^\top &= E\left[\frac{\partial}{\partial \btheta_R} U(a,\bfO_i; \btheta_R,\btheta_Y^*, \btheta_A^*)\Big|_{\btheta_R=\btheta_R^*}\right]^\top E\left[\frac{\partial}{\partial \btheta_R}\bpsi_R(\bfO_i;\btheta_R)\Big|_{\btheta_R=\btheta_R^*}\right], \\
            \boldsymbol{q}_{2a}^\top &= E\left[\frac{\partial}{\partial \btheta_Y} U(a,\bfO_i; \btheta_R^*,\btheta_Y, \btheta_A^*)\Big|_{\btheta_Y=\btheta_Y^*}\right]^\top E\left[\frac{\partial}{\partial \btheta_Y}\bpsi_Y(\bfO_i;\btheta_Y)\Big|_{\btheta_Y=\btheta_Y^*}\right],\\
            \boldsymbol{q}_{3a}^\top &= E\left[\frac{\partial}{\partial \btheta_Y} U(a,\bfO_i; \btheta_R^*,\btheta_Y^*, \btheta_A)\Big|_{\btheta_A=\btheta_A^*}\right]^\top E\left[\frac{\partial}{\partial \btheta_Y}\bpsi_A(\bfO_i;\btheta_A)\Big|_{\btheta_Y=\btheta_Y^*}\right].
        \end{align*}
        Following the proof of Proposition 2, $\boldsymbol{q}_{2a} = \bzero$ since $\pi(\bfB_i;\btheta_A)$ and $\kappa_{ij}(a,\btheta_R)$ are correct specified. 
        Therefore, denoting
        \begin{align*}
        \bU(\bfO_i;\btheta_R,\btheta_{Y}, \btheta_A) &= (U(1,\bfO_i;\btheta_R,\btheta_Y, \btheta_A), U(0,\bfO_i;\btheta_R,\btheta_Y, \btheta_A))^\top,\\
        \mathbf{q}_1 &= (\bq_{11},\bq_{10}),\\
        \mathbf{q}_3 &= (\bq_{31},\bq_{30}),
        \end{align*}
        the asymptotic variance of $(\widehat{\mu}_1, \widehat{\mu}_0)$ is
        \begin{align*}
            \bfV_{\mu} &= Var\{\bU(\bfO_i;\btheta_R^*,\btheta_{Y}^*, \btheta_A^*) - \mathbf{q}_1^\top \bpsi_R(\bfO_i;\btheta_R^*) - \mathbf{q}_3^\top \bpsi_A(\bfO_i;\btheta_A^*)\} \\
            &= \underbrace{Var\{\bU(\bfO_i;\btheta_R^*,\btheta_{Y}^*, \btheta_A^*) - \mathbf{q}_1^\top \bpsi_R(\bfO_i;\btheta_R^*)\}}_{\textup{denoted as (a)}} + \underbrace{Var\{\mathbf{q}_3^\top \bpsi_A(\bfO_i;\btheta_A^*)\}}_{\textup{denoted as (b)}}\\
             &\quad  \underbrace{-2Cov\{\bU(\bfO_i,\btheta_R^*,\btheta_{Y}^*, \btheta_A^*), \mathbf{q}_3^\top \bpsi_A(\bfO_i;\btheta_A^*)\}}_{\textup{denoted as (c)}} + \underbrace{2Cov\{\mathbf{q}_1^\top \bpsi_R(\bfO_i;\btheta_R^*),\mathbf{q}_3^\top \bpsi_A(\bfO_i;\btheta_A^*)\}}_{\textup{denoted as (d)}}.
        \end{align*}
        In the above expression of $\bfV_{\mu}$, (a) is the asymptotic variance when the treatment assignment is not modeled (Equation~\eqref{def: v_mu_prop2}). Therefore, to prove that modeling treatment assignment does not increase variance, we just need to show (b)+(c)+(d) is negative semi-definite.
        To this end, we first show (d) = 0. Since we assume that the outcome missing model is correctly specified in the sense that $E[\bpsi_R(\bfO_i;\btheta_R^*)|A_i,\bfB_i] = \bzero$,  we have
        \begin{align*}
            E\left[\bpsi_A(\bfO_i;\btheta_A^*)\bpsi_R(\bfO_i;\btheta_R^*)^\top\right]= E[\bpsi_A(\bfO_i;\btheta_A^*)E[\bpsi_R(\bfO_i;\btheta_R^*)|A_i,\bfB_i]^\top] = \bzero,
        \end{align*}
        which implies (d) = 0. Next, we show (c)=$-$2(b). Since (b) is positive semi-definite, then (c)=$-$2(b) yields (b)+(c)+(d) is negative semi-definite, which will complete the proof. To prove (c)=$-$2(b), we first observe that, because we use maximum likelihood estimation for $\btheta_A$, we have $$\bpsi_A(\bfO_i;\btheta_A) = \frac{A_i-\pi(\bfB_i;\btheta_A)}{\pi(\bfB_i;\btheta_A)\{1-\pi(\bfB_i;\btheta_A)\}}  \pi'(\bfB_i;\btheta_A),$$
        where $\pi'(\bfB_i;\btheta_A)$ is the partial derivative of $\pi(\bfB_i;\btheta_A)$ on $\btheta_A$ evaluated at $\btheta_A$. This result implies
        \begin{align*}
            &E\left[\frac{\partial }{\partial \btheta_A} \bpsi_A(\bfO_i; \btheta_A) \big |_{\btheta_A = \btheta_A^*}\right] = -E\left[\frac{\pi'(\bfB_i,\btheta_A^*)\pi'(\bfB_i,\btheta_A^*)^\top}{\pi(1-\pi)}\right] =-E\left[\bpsi_A(\bfO_i;\btheta_A^*) \bpsi_A(\bfO_i;\btheta_A^*)^\top\right].
        \end{align*}
 Furthermore, direct algebra shows that
        \begin{align*}
            E\left[\frac{\partial }{\partial \btheta_A} U(a, \bfO_i; \btheta_R^*,\btheta_Y^*, \btheta_A) \big |_{\btheta_A = \btheta_A^*}\right] &= -E\left[\sum_{j=1}^{N_i} w_{ij}^*\frac{(-1)^{1-a}\pi'(\bfB_i,\btheta_A^*)}{\pi^{a}(1-\pi)^{1-a}} \left\{E[Y_{ij}|A_i=a,\bfB_i] - \eta_{ij}(a,\btheta_Y^*)\right\}\right] \\
            &= - E\left[\bpsi_A(\bfO_i;\btheta_A^*) \{U(a,\bfO_i;\btheta_R^*,\btheta_Y^*, \btheta_A^*) -  \mu_1^*\}\right].
        \end{align*}
Putting together, we obtain
\begin{align*}
    \bq_{3a} &= E\left[\frac{\partial }{\partial \btheta_A} \bpsi_A(\bfO_i; \btheta_A) \big |_{\btheta_A = \btheta_A^*}\right]^{-1}E\left[\frac{\partial }{\partial \btheta_A} U(a, \bfO_i; \btheta_R^*,\btheta_Y^*, \btheta_A) \big |_{\btheta_A = \btheta_A^*}\right] \\
    &= E\left[\bpsi_A(\bfO_i;\btheta_A^*) \bpsi_A(\bfO_i;\btheta_A^*)^\top\right]^{-1}E\left[\bpsi_A(\bfO_i;\btheta_A^*) \{U(a,\bfO_i;\btheta_R^*,\btheta_Y^*, \btheta_A^*) -  \mu_1^*\}\right] \\
    &= Var\{\bpsi_A(\bfO_i;\btheta_A^*)\}^{-1} Cov\{\bpsi_A(\bfO_i;\btheta_A^*),U(a,\bfO_i;\btheta_R^*,\btheta_Y^*, \btheta_A^*)\}.
\end{align*}
As a result 
\begin{align*}
    \textup{(c)} = -2 Cov\{\bU(\bfO_i,\btheta_R^*,\btheta_{Y}^*, \btheta_A^*), \bpsi_A(\bfO_i;\btheta_A^*)\}\mathbf{q}_3 = -2 \mathbf{q}_3^\top  Var\{\bpsi_A(\bfO_i;\btheta_A^*)\}^{-1}\mathbf{q}_3 = -2\textup{(b)},
\end{align*}
which completes the proof that modeling treatment assignment does not increase variance.
Finally, if the outcome model is correctly specified,  then $\bq_{3a} = \bzero$, meaning that the model $\pi(\bfB_i;\btheta_A)$ does not modify the variance.
\end{proof}

\subsection{Proof of Theorem 3}
\begin{lemma}
    Given Assumptions 1, 2, and 6, 
    \begin{align*}
        E[Y_{ij}|R_{ij}^Y=1,E_{ij}=1, A_i,\bB_{ij}] &= E[Y_{ij}|R_{ij}^Y=1, A_i,\bB_{ij}] \\
        E[Y_{ij}|E_{ij}=1, A_i,\bB_{ij}] &= E[Y_{ij}|A_i,\bB_{ij}] \\
        E[R_{ij}^Y|E_{ij}=1, A_i,\bB_{ij}] &= E[R_{ij}^Y|A_i,\bB_{ij}]
    \end{align*}
\end{lemma}
\begin{proof}
    Let $\bB_{ij}^{\textup{(1)}}$ be $\bB_{ij}$ with $R_i^N$ set to be 1 and $\bB_{ij}^{\textup{(0)}}$ be $\bB_{ij}$ with $R_i^N$ set to be 0. Then
    \begin{align*}
        &E[Y_{ij}|R_{ij}^Y=1,E_{ij}=1, A_i,\bB_{ij}] \\
        &= I\{R_i^N =1\}E[Y_{ij}|R_{ij}^Y=1,E_{ij}=1, A_i,\bB_{ij}^{\textup{(1)}}] + I\{R_i^N =0\}E[Y_{ij}|R_{ij}^Y=1,E_{ij}=1, A_i,\bB_{ij}^{\textup{(0)}}].
    \end{align*}
    For $E[Y_{ij}|R_{ij}^Y=1,E_{ij}=1, A_i,\bB_{ij}^{\textup{(1)}}]$, since $(M_i,N_i)$ is part of $\bB_{ij}^{\textup{(1)}}$, then Assumption 6 (1) implies that $E[Y_{ij}|R_{ij}^Y=1,E_{ij}=1, A_i,\bB_{ij}^{\textup{(1)}}] = E[Y_{ij}|R_{ij}^Y=1, A_i,\bB_{ij}^{\textup{(1)}}]$. For the term with $\bB_{ij}^{\textup{(0)}}$, we have
    \begin{align*}
        & E[Y_{ij}|R_{ij}^Y=1,E_{ij}=1, A_i,\bB_{ij}^{\textup{(0)}}]\\
        &=E\left[ E[Y_{ij}|R_{ij}^Y=1,E_{ij}=1, A_i,\bB_{ij}^{\textup{(0)}}, N_i]\Big|R_{ij}^Y=1,E_{ij}=1, A_i,\bB_{ij}^{\textup{(0)}}\right] \\
        &=E\left[ E[Y_{ij}|R_{ij}^Y=1,A_i,\bB_{ij}^{\textup{(0)}}, N_i]\Big|R_{ij}^Y=1,E_{ij}=1, A_i,\bB_{ij}^{\textup{(0)}}\right] \\
        &=E\left[ E[Y_{ij}|R_{ij}^Y=1,A_i,\bB_{ij}^{\textup{(0)}} ]\Big|R_{ij}^Y=1,E_{ij}=1, A_i,\bB_{ij}^{\textup{(0)}}\right] \\
        &=E[Y_{ij}|R_{ij}^Y=1,A_i,\bB_{ij}^{\textup{(0)}} ],
    \end{align*}
    where the second equation results from Assumption 6 (1), the third equation results from Assumption 6 (2), and the last equation results from the definition of conditional expectation. Therefore,
        \begin{align*}
        &E[Y_{ij}|R_{ij}^Y=1,E_{ij}=1, A_i,\bB_{ij}] \\
        &= I\{R_i^N =1\}E[Y_{ij}|R_{ij}^Y=1,E_{ij}=1, A_i,\bB_{ij}^{\textup{(1)}}] + I\{R_i^N =0\}E[Y_{ij}|R_{ij}^Y=1,E_{ij}=1, A_i,\bB_{ij}^{\textup{(0)}}]\\
        &= I\{R_i^N =1\}E[Y_{ij}|R_{ij}^Y=1,A_i,\bB_{ij}^{\textup{(1)}}] + I\{R_i^N =0\}E[Y_{ij}|R_{ij}^Y=1,A_i,\bB_{ij}^{\textup{(0)}}] \\
        &=E[Y_{ij}|R_{ij}^Y=1,A_i,\bB_{ij}].
    \end{align*}
    Then, we can follow a similar proof to obtain $E[Y_{ij}|E_{ij}=1, A_i,\bB_{ij}] = E[Y_{ij}|A_i,\bB_{ij}]$ and $ E[R_{ij}^Y|E_{ij}=1, A_i,\bB_{ij}] = E[R_{ij}^Y|A_i,\bB_{ij}]$.
\end{proof}

\begin{proof}[Proof of Theorem 3.]
By the regularity conditions, we can use the exact same proof for Theorem 1 to obtain $\widehat{\btheta} \xrightarrow{P} \btheta^*$ and asymptotic normality. Then, we only need to show when $\mu_a^* = E[Y_{ij}|A_i=a]$, i.e., double robustness. To this end, we define
\begin{align*}
    H_{ij} =  \frac{I\{A_i=a\}}{\pi^a(1-\pi)^{1-a}}     \frac{R_{ij}^Y \{Y_{ij} - \eta(a, \bB_{ij};\btheta_Y^*)\}}{ \kappa(a, \bB_{ij};\btheta_R^*)} + \eta(a,\bB_{ij};\btheta_Y^*),
\end{align*}    
which is a function of $(A_i, R_{ij}^Y, Y_{ij}, \bB_{ij})$. Then
\begin{align*}
    \mu_a^* &= E\left[\frac{1}{M_i}\sum_{j=1}^{N_i} E_{ij} H_{ij}\right] \\
    &= E\left[\frac{1}{M_i}\sum_{j=1}^{N_i} E[E_{ij} H_{ij}|\bR_i^C, \bR_i^C \circ \bC_i, R_i^N, N_i, M_i]\right] \\
    &= E\left[\frac{1}{M_i}\sum_{j=1}^{N_i} E[E_{ij}|H_{ij}, \bR_i^C, \bR_i^C \circ \bC_i, R_i^N, N_i, M_i] E[ H_{ij}|\bR_i^C, \bR_i^C \circ \bC_i, R_i^N, N_i, M_i]\right] \\
    &= E\left[\frac{1}{M_i}\sum_{j=1}^{N_i} \frac{M_i}{N_i} E[ H_{ij}|\bR_i^C, \bR_i^C \circ \bC_i, R_i^N, N_i, M_i]\right] \\
    &=E\left[\frac{1}{N_i}\sum_{j=1}^{N_i} E[ H_{ij}|\bR_i^C, \bR_i^C \circ \bC_i, R_i^N, N_i, M_i]\right],
\end{align*}
where the fourth equation uses Assumption 6 (1).
The above derivation leads us to study the quantity $E[ H_{ij}|\bR_i^C, \bR_i^C \circ \bC_i, R_i^N, N_i, M_i]$.
By Assumption 6 (2), $E[ H_{ij}|\bR_i^C, \bR_i^C \circ \bC_i, R_i^N=0, N_i, M_i] = E[ H_{ij}|\bR_i^C, \bR_i^C \circ \bC_i, R_i^N=0,  M_i]$, which implies
\begin{align*}
    E[ H_{ij}|\bR_i^C, \bR_i^C \circ \bC_i, R_i^N, N_i, M_i] &= E[ H_{ij}|\bR_i^C, \bR_i^C \circ \bC_i, R_i^N, R_i^NN_i, M_i] \\
    &=E[ E[H_{ij}|\bB_{ij}]|\bR_i^C, \bR_i^C \circ \bC_i, R_i^N, R_i^NN_i, M_i]
\end{align*}
since $(\bR_i^C, \bR_i^C \circ \bC_i, R_i^N, R_i^NN_i, M_i)$ is part of the observed covariate data $\bB_{ij}$. To compute $E[H_{ij}|\bB_{ij}]$, we recall that Assumption 2 implies $P(A=a|\bB_{ij}) = \pi^a(1-\pi)^{1-a}$ and get
\begin{align*}
    E[H_{ij}|\bB_{ij}] &=      \frac{E[R_{ij}^Y|A_i=a,\bB_{ij}] \{E[Y_{ij}|R_{ij}^Y=1,A_i=a,\bB_{ij}] - \eta(a, \bB_{ij};\btheta_Y^*)\}}{ \kappa(a, \bB_{ij};\btheta_R^*)}+ \eta(a,\bB_{ij};\btheta_Y^*).
\end{align*}
When $\eta(a, \bB_{ij};\btheta_Y^*) = E[Y_{ij}|R_{ij}^Y=1,E_{ij}=1,A_i=a,\bB_{ij}]$ or $\kappa(a, \bB_{ij};\btheta_R^*) = E[R_{ij}^Y|A_i=a,E_{ij}=1,\bB_{ij}]$, Lemma 1 implies that $E[H_{ij}|\bB_{ij}] = E[Y_{ij}|R_{ij}^Y=1,E_{ij}=1,A_i=a,\bB_{ij}]$. By Assumption 6 (3) and Lemma~1, $E[Y_{ij}|R_{ij}^Y=1,E_{ij}=1,A_i=a,\bB_{ij}] = E[Y_{ij}|A_i=a,\bB_{ij}]$.
Finally, we have
\begin{align*}
    \mu_a^* &=E\left[\frac{1}{N_i}\sum_{j=1}^{N_i} E[ E[H_{ij}|\bB_{ij}]|\bR_i^C, \bR_i^C \circ \bC_i, R_i^N, R_i^NN_i, M_i]\right] \\
    &= E\left[\frac{1}{N_i}\sum_{j=1}^{N_i} E[ E[Y_{ij}|A_i=a,\bB_{ij}]|\bR_i^C, \bR_i^C \circ \bC_i, R_i^N, R_i^NN_i, M_i]\right] \\
    &=E\left[\frac{1}{N_i}\sum_{j=1}^{N_i} E[ Y_{ij}|A_i=a,\bR_i^C, \bR_i^C \circ \bC_i, R_i^N, R_i^NN_i, M_i]\right] \\
    &= E\left[ E[ Y_{ij}|A_i=a,\bR_i^C, \bR_i^C \circ \bC_i, R_i^N, R_i^NN_i, M_i]\right] \\
    &= E[ Y_{ij}|A_i=a],
\end{align*}
where the second last equation comes from Assumption 1 (2), and the last equation comes from Assumption 2. 
\end{proof}

In this scenario with subsampling, to obtain similar results as in Propositions A-C, we need the assumption that $(R_{ij}, Y_{ij})$ is independent of $\bB_{ij'}$ given $A_i, \bB_{ij}$ $j\ne j'$. In this case, we have $E[R_{ij}|A_i, \bfB_i] = E[R_{ij}|A_i, \bB_{ij}]$ and $E[R_{ij}Y_{ij}|A_i, \bfB_i] = E[R_{ij}Y_{ij}|A_i, \bB_{ij}]$. Therefore, all derivations remain the same.

\subsection{Proof of Proposition 2 of the main paper}
\begin{proof}
Recall the definition
\begin{align*}
    H_{ij} =  \frac{I\{A_i=a\}}{\pi^a(1-\pi)^{1-a}}     \frac{R_{ij}^Y \{Y_{ij} - \eta(a, \bB_{ij};\btheta_Y^*)\}}{ \kappa(a, \bB_{ij};\btheta_R^*)} + \eta(a,\bB_{ij};\btheta_Y^*).
\end{align*}
Under Assumption (1) and regularity conditions, we obtain that
\begin{align*}
    \underline{\mu}_a &=  E\left[\frac{1}{M}\sum_{j=1}^{N_i} E_{ij}H_{ij}\right] \\
    &= E\left[\frac{1}{M}\sum_{j=1}^{N_i}E[E_{ij}H_{ij}|M_i,N_i]\right] \\
    &= E\left[\frac{1}{M}\sum_{j=1}^{N_i}E[E_{ij}|M_i,N_i]E[H_{ij}|E_{ij}=1,M_i,N_i]\right].
\end{align*}
Since we assume that $E_{ij}$ is identically distributed given $(M_i, N_i)$, we can show that $E[E_{ij}|N_i]=M_i/N_i$. To see this, by definition, $\sum_{j=1}^{N_i}E[E_{ij}|M_i,N_i] = E[\sum_{j=1}^{N_i} E_{ij}|M_i,N_i] = E[M_i|M_i,N_i] = M_i$. On the other hand, since that $E[E_{ij}|M_i,N_i]$ does not vary across $j$, then $\sum_{j=1}^{N_i}E[E_{ij}|M_i,N_i] = N_i E[E_{ij}|M_i,N_i]$. Combined together, we get $E[E_{ij}|M_i,N_i] = M_i/N_i$ for all $j$. 

Furthermore, since we assume $\{(R_{ij}^Y, Y_{ij}, E_{ij}, \bB_{ij}),j=1,\dots, N_i\}$ are identically distributed given $N_i$, then $E[H_{ij}|E_{ij}=1,M_i,N_i]$ does not vary across $j$, which implies
\begin{align*}
    \underline{\mu}_a = E[E[H_{ij}|E_{ij}=1,M_i,N_i]].
\end{align*}
To further simplify the above expression, since $(M_i, N_i)$ are contained in $\bB_{ij}$, we have
\begin{align*}
    & E[H_{ij}|E_{ij}=1,M_i,N_i]\\
    &= E[E[H_{ij}|E_{ij}=1,\bB_{ij}]|E_{ij}=1,M_i,N_i] \\
    &= E\left[\frac{P(A_i=a|E_{ij}=1,\bB_{ij})}{\pi^a(1-\pi)^{1-a}}\frac{P(R_{ij}^Y=1|E_{ij}=1,A_i=a, \bB_{ij})}{\kappa(a, \bB_{ij};\btheta_R^*)}\right.\\
    &\quad \left. \times \{E[Y_{ij}|R_{ij}^Y=1,E_{ij}=1,A_i=a, \bB_{ij}] - \eta(a, \bB_{ij};\btheta_Y^*)\} + \eta(a, \bB_{ij};\btheta_Y^*)\bigg|E_{ij}=1,M_i,N_i \right]
\end{align*}
By the Assumption (2), $P(A_i=a|E_{ij}=1,\bB_{ij}) = \pi^a(1-\pi)^{1-a}$. When $P(R_{ij}^Y=1|E_{ij}=1,A_i=a, \bB_{ij}) = \kappa(a, \bB_{ij};\btheta_R^*)$ or $E[Y_{ij}|R_{ij}^Y=1,E_{ij}=1,A_i=a, \bB_{ij}] = \eta(a, \bB_{ij};\btheta_Y^*)$, then we get
\begin{align*}
    E[H_{ij}|E_{ij}=1,M_i,N_i] = E[E[Y_{ij}|R_{ij}^Y=1,E_{ij}=1,A_i=a, \bB_{ij}]|E_{ij}=1,M_i,N_i].
\end{align*}
Hence,
$\underline{\mu}_a = E[E[E[Y_{ij}|R_{ij}^Y=1,E_{ij}=1,A_i=a, \bB_{ij}]|E_{ij}=1,M_i,N_i]]$.

To derive the difference between $\underline{\mu}_a$ and $E[Y_{ij}|A_i=a]$ we observe that
\begin{align*}
    E[Y_{ij}|A_i=a] &= E[E[Y_{ij}|A_i=a, M_i, N_i]] \\
    &= E[E[E[Y_{ij}|E_{ij},A_i=a, M_i, N_i]| M_i, N_i]]\\
    &= E\left[\sum_{e=0}^1 E[Y_{ij}|E_{ij}=e,A_i=a, M_i, N_i]P(E_{ij}=e|M_i, N_i)\right] \\
    &= E[E[Y_{ij}|E_{ij}=1,A_i=a, M_i, N_i]] - E\left[\frac{N_i-M_i}{N_i}\delta_{a}(M_i, N_i)\right],
\end{align*}
where the second equation results from Assumption (2). 
For the first term  $E[E[Y_{ij}|E_{ij}=1,A_i=a, M_i, N_i]]$ above, we have
\begin{align*}
    & E[Y_{ij}|E_{ij}=1,A_i=a, M_i, N_i]\\
    &= E[E[Y_{ij}|E_{ij}=1,A_i=a, \bB_{ij}]|E_{ij}=1,M_i, N_i] \\
    &= E[E[E[Y_{ij}|R_{ij}^Y,E_{ij}=1,A_i=a, \bB_{ij}]|E_{ij}=1,A_i=a,\bB_{ij}]|E_{ij}=1,M_i, N_i] \\
    &= E\left[\sum_{r=0}^1 E[Y_{ij}|R_{ij}^Y=r,E_{ij}=1,A_i=a, \bB_{ij}]P(R_{ij}^Y=r|E_{ij}=1,A_i=a, \bB_{ij})\bigg|E_{ij}=1,M_i, N_i\right] \\
    &= E[E[Y_{ij}|R_{ij}^Y=1,E_{ij}=1,A_i=a, \bB_{ij}]|E_{ij}=1,M_i, N_i] \\
    &\quad - E[ \gamma_{a}(\bB_{ij})P(R_{ij}^Y=0|E_{ij}=1,A_i=a, \bB_{ij})|E_{ij}=1,M_i, N_i] \\
    &=E[E[Y_{ij}|R_{ij}^Y=1,E_{ij}=1,A_i=a, \bB_{ij}]|E_{ij}=1,M_i, N_i]\\
    &\quad - E[(1-R_{ij}^Y)\gamma_{a}(\bB_{ij})|E_{ij}=1,A_i=a, M_i, N_i].
\end{align*}
Therefore, 
\begin{align*}
    \underline{\mu}_a - E[Y_{ij}|A_{ij}=a] = E\left[\frac{N_i-M_i}{N_i}\delta_{a}(M_i, N_i)\right] + E[E[(1-R_{ij}^Y)\gamma_{a}(\bB_{ij})|E_{ij}=1,A_i=a, M_i, N_i]].
\end{align*}
\end{proof}

\section{Additional simulation studies}\label{asec: simulation}
\subsection{Simulation with 10 clusters}
The simulation setting for $m=10$ is the same as the first simulation in the main paper, and we made an additional change to only adjust for the covariate $X_{ij1}^*$ in all regression models. When $m=30$ or $100$, we adjusted for all 7 covariates, which is infeasible for 10 clusters, especially after adjustment for degrees of freedom. 
    
    The simulation results are replicated in Table~\ref{tab:sim1-m10} below.
    The proposed estimators still have negligible bias and near 95\% coverage probability, while we see 1-3\% undercoverage due to small samples, as expected. These results further support the small-sample applicability of our methods. 
    For the unadjusted, IPW, and DR estimators, the bias remains similar to the setting with $m=100$, and the standard errors all increase due to the reduced sample sizes. As a result, increased variance mitigates the undercoverage from bias, while 3-12\% undercoverage remains. 
    \begin{table}[htbp]
    \centering
    \caption{Results for the first simulation study with $m=10$. ESE: empirical standard error. ASE: average of standard error estimators. CP: coverage probability of 95\% confidence intervals based on $t$-distributions with degree of freedom $m-q$, where $q$ is the number of covariates used in the analysis. }
    \label{tab:sim1-m10}
    \begin{tabular}{ccrrrrrcrrrr}
    \hline
          \multirow{2}{*}{Sampling}  &\multirow{2}{*}{$m$} & \multirow{2}{*}{Estimator}&   \multicolumn{4}{c}{10\% missing data}& &\multicolumn{4}{c}{30\% missing data}\\
          \cline{4-7} \cline{9-12}
           & & & 
     Bias& ESE& ASE& CP&& Bias& ESE& ASE& CP\\
    \hline
  \multirow{5}{*}{No}&\multirow{5}{*}{10}& Unadjusted&  -0.32 & 2.01 & 1.96 & 0.92&& -0.97 & 1.92 & 1.89 & 0.87 \\ 
  && IPW& 0.90 & 2.50 & 1.77 & 0.83&&  0.92 & 2.45 & 1.79 & 0.84 \\ 
  && DR & 1.01 & 2.07 & 1.70 & 0.88&& 1.00 & 2.09 & 1.77 & 0.90 \\ 
  && DR-PM& -0.08 & 1.70 & 1.65 & 0.92&& -0.06 & 1.76 & 1.72 & 0.93 \\ 
  && DR-ML& 0.00 & 1.62 & 1.60 & 0.94&& -0.02 & 1.72 & 1.67 & 0.94 \\ 
  \hline
  \multirow{5}{*}{Yes}&\multirow{5}{*}{10}& Unadjusted& -0.24 & 2.04 & 1.96 & 0.92&& -1.05 & 2.01 & 1.89 & 0.85 \\  
  && IPW& 1.01 & 2.41 & 1.77 & 0.83&&  0.78 & 2.59 & 1.77 & 0.82 \\ 
  && DR & 1.10 & 2.01 & 1.69 & 0.91&& 0.90 & 2.19 & 1.74 & 0.87 \\ 
  && DR-PM& -0.01 & 1.73 & 1.70 & 0.93&& -0.17 & 1.86 & 1.76 & 0.92 \\ 
  && DR-ML&  0.04 & 1.65 & 1.62 & 0.94&& -0.12 & 1.83 & 1.68 & 0.92 \\ 
    \hline
    \end{tabular}
\end{table}
\subsection{Simulation with no treatment effects}

We repeat the first simulation study setting $\Delta=0$ by subtracting a constant treatment effect from the outcome model. The results are presented in Table~\ref{tab:sim1-null}, which are similar to the simulation results in the main paper. This is expected because the theoretical properties of our proposed estimators are not affected by the magnitude of the true treatment effect. 

\begin{table}[htbp]
    \centering
    \caption{Results for the first simulation study with no treatment effect. ESE: empirical standard error. ASE: average of standard error estimators. CP: coverage probability of 95\% confidence intervals based on $t$-distributions with degree of freedom $m-q$, where $q$ is the number of covariates used in the analysis. }
    \label{tab:sim1-null}
    \begin{tabular}{ccrrrrrcrrrr}
    \hline
          \multirow{2}{*}{Sampling}  &\multirow{2}{*}{$m$} & \multirow{2}{*}{Estimator}&   \multicolumn{4}{c}{10\% missing data}& &\multicolumn{4}{c}{30\% missing data}\\
          \cline{4-7} \cline{9-12}
           & & & 
     Bias& ESE& ASE& CP&& Bias& ESE& ASE& CP\\
    \hline
  \multirow{10}{*}{No}&\multirow{5}{*}{30}& Unadjusted& -0.35 & 1.16 & 1.11 & 0.92&& -1.02 & 1.07 & 1.08 & 0.83 \\ 
  && IPW& 1.03 & 1.41 & 1.26 & 0.86&& 1.01 & 1.34 & 1.28 & 0.88 \\ 
  && DR & 1.06 & 1.19 & 1.06 & 0.84&& 1.05 & 1.18 & 1.06 & 0.85 \\
  && DR-PM& 0.01 & 1.13 & 1.08 & 0.94&& -0.04 & 1.14 & 1.09 & 0.92 \\
  && DR-ML& 0.01 & 0.83 & 0.77 & 0.92&& -0.02 & 0.89 & 0.80 & 0.92 \\ 
  \cline{2-12}
  &\multirow{5}{*}{100}& Unadjusted& -0.35 & 0.58 & 0.61 & 0.92&& -0.97 & 0.56 & 0.59 & 0.62 \\ 
  && IPW& 1.07 & 0.71 & 0.72 & 0.70&& 1.10 & 0.71 & 0.73 & 0.69 \\ 
  && DR & 1.08 & 0.58 & 0.58 & 0.55&& 1.11 & 0.58 & 0.58 & 0.53 \\ 
  && DR-PM& -0.01 & 0.49 & 0.51 & 0.97&& 0.02 & 0.48 & 0.52 & 0.97 \\ 
  && DR-ML& -0.02 & 0.40 & 0.37 & 0.92&& 0.00 & 0.41 & 0.40 & 0.95 \\ 
  \hline
  \multirow{10}{*}{Yes}&\multirow{5}{*}{30}& Unadjusted& -0.36 & 1.15 & 1.13 & 0.93&& -0.98 & 1.14 & 1.10 & 0.83 \\ 
  && IPW& 1.06 & 1.39 & 1.27 & 0.86&&  1.06 & 1.41 & 1.29 & 0.84 \\
  && DR & 1.09 & 1.20 & 1.08 & 0.84&& 1.06 & 1.19 & 1.08 & 0.85 \\
  && DR-PM& 0.03 & 1.11 & 1.07 & 0.95&& 0.00 & 1.37 & 1.18 & 0.93 \\ 
  && DR-ML& 0.03 & 0.87 & 0.80 & 0.94&& -0.04 & 0.90 & 0.83 & 0.92 \\ 
  \cline{2-12}
  &\multirow{5}{*}{100}& Unadjusted& -0.36 & 0.60 & 0.61 & 0.91&& -0.99 & 0.59 & 0.60 & 0.64 \\ 
  && IPW& 1.04 & 0.73 & 0.73 & 0.72 && 1.11 & 0.74 & 0.74 & 0.70 \\ 
  && DR & 1.06 & 0.60 & 0.58 & 0.57&& 1.13 & 0.61 & 0.59 & 0.52 \\
  && DR-PM& -0.01 & 0.49 & 0.52 & 0.96&& 0.01 & 0.52 & 0.54 & 0.96 \\ 
  && DR-ML& -0.03 & 0.40 & 0.38 & 0.93 && 0.00 & 0.44 & 0.41 & 0.93 \\ 
    \hline
    \end{tabular}
\end{table}

\subsection{Comparison with two-stage TMLE}
To evaluate the performance of Two-stage TMLE \citep{balzer2023two}, we considered the setting of the first simulation study. We followed the Supplementary Material of \citep{balzer2023two} to implement this approach and used the R package, `tmle' (Version 2.0.1.1, updated on 2024 May, \citealp{tmle-package}). At Stage 1 and 2 of this method, we imputed missing covariates by zero, adjusted for all missing covariate indicators and all imputed covariates in nuisance function estimation (including outcome regression at both stages, outcome missingness at Stage 1, and treatment assignment at Stage 2), adopted the ensemble learner of mean, generalized linear models, and generalized additive models (same as  in \citealp{balzer2023two}) with 2-fold validation for stability (necessary when $M_i$ is small). We did not implement adaptive pre-specification for variable selection; this choice does not affect consistency but may cause variance increase.

The results of two-stage TMLE are presented in Table~\ref{tab:sim1-tmle}. Overall, Two-stage TMLE performs similarly to CDR-PM, while it has moderate bias when there are 30\% missing data and within-cluster sampling. Theoretically, two-stage TMLE is a consistent estimator in our simulation setting, but this bias occurs because a larger proportion of missing outcomes leads to a smaller effective sample size per cluster (sometimes less than 10 data points), which causes challenges at Stage 1, where TMLE models are fit within each cluster. 
(For example, in a cluster with 3 individuals and 2 missing outcomes, it is computationally infeasible to implement TMLE.)  
In fact, when $p_m=0.3$, two-stage TMLE returns an error  in more than 50\% data replicates. After we exclude all problematic simulations, it is not surprising the remaining results show some bias. In practice, this issue can be avoided when missing data are light or all clusters have large sample sizes. In terms of variance, two-stage TMLE is less precise than CDR-ML in this example, because the latter adopts regression trees in the ensemble learner---regression trees can better model outcomes in our data generating distribution but may cause over-fitting in other scenarios. The precision of two-stage TMLE can be substaintially improved by using regression trees and adaptive pre-specification. Finally, in terms of computation time, two-stage TMLE requires model fitting within each cluster and takes $\sim10$ minutes for one run (on Macbook Pro 2024 with M3 chip). In contrast, all other five methods take less than 1 second. This difference in computation time can be negligible in most real-wolrd applications.

\begin{table}[htbp]
    \centering
    \caption{Results for the first simulation study. ESE: empirical standard error. ASE: average of standard error estimators. CP: coverage probability of 95\% confidence intervals based on $t$-distributions. }
    \label{tab:sim1-tmle}
    \begin{tabular}{ccrrrrrcrrrr}
    \hline
          \multirow{2}{*}{Sampling}  &\multirow{2}{*}{$m$} & \multirow{2}{*}{Estimator}&   \multicolumn{4}{c}{10\% missing data}& &\multicolumn{4}{c}{30\% missing data}\\
          \cline{4-7} \cline{9-12}
           & & & 
     Bias& ESE& ASE& CP&& Bias& ESE& ASE& CP\\
    \hline
  \multirow{6}{*}{No}&\multirow{3}{*}{30}& Two-stage TMLE & -0.01 & 1.12 & 1.07 & 0.92&& -0.21 & 1.11 & 1.04 & 0.91 \\ 
  && CDR-PM& -0.00 & 1.13 & 1.08 & 0.94&& -0.02 & 1.14 & 1.08 & 0.92 \\
  && CDR-ML& 0.00 & 0.82 & 0.77 & 0.93&& -0.01 & 0.89 & 0.80 & 0.92 \\ 
  \cline{2-12}
  &\multirow{3}{*}{100}& Two-stage TMLE & -0.04 & 0.51 & 0.49 & 0.94 && -0.18 & 0.52 & 0.50 & 0.92 \\ 
  && CDR-PM& -0.01 & 0.49 & 0.51 & 0.96&& 0.02 & 0.50 & 0.52 & 0.96 \\ 
  && CDR-ML& -0.02 & 0.40 & 0.37 & 0.92&& -0.00 & 0.42 & 0.40 & 0.94 \\ 
  \hline
  \multirow{6}{*}{Yes}&\multirow{3}{*}{30}& Two-stage TMLE& -0.02 & 1.23 & 1.08 & 0.89 && -0.39 & 1.11 & 1.05 & 0.90 \\ 
  && CDR-PM& 0.05 & 1.13 & 1.07 & 0.95&& -0.02 & 1.16 & 1.17 & 0.95 \\ 
  && CDR-ML& 0.02 & 0.86 & 0.79 & 0.94&& -0.02 & 0.92 & 0.83 & 0.92 \\ 
  \cline{2-12}
  &\multirow{3}{*}{100}& Two-stage TMLE& 0.01 & 0.50 & 0.51 & 0.94&& -0.35 & 0.50 & 0.52 & 0.91 \\ 
  && CDR-PM& -0.01 & 0.50 & 0.52 & 0.96&& 0.00 & 0.52 & 0.54 & 0.96 \\ 
  && CDR-ML& -0.03 & 0.40 & 0.38 & 0.94 && -0.02 & 0.44 & 0.41 & 0.93 \\ 
    \hline
    \end{tabular}
\end{table}

\section{Sensitivity analysis for the data application}\label{asec: sensitivity}
For the WFHS, we additionally performed a sensitivity analysis for the missing data assumptions. As in the simulation studies, we adopt the constant sensitivity parameters: $\delta_1-\delta_0$ and $\gamma_1 -\gamma_0$. Under the assumption of no selection bias, i.e., $\delta_1-\delta_0 = 0$, the tipping point for DR-ML is $\gamma_1-\gamma_0 = 0.46$. Here, $\gamma_1$ and $\gamma_0$ denote the treatment effects among observed and unobserved outcomes, respectively. Using the unadjusted estimator as $\widehat{\gamma}_1 = 0.17$ from Table 4 of the main paper, the intervention remains effective as long as $\gamma_0 \ge -0.29$, i.e., the treatment effect for individuals with missing outcomes would need to be no smaller than $-0.29$. Given the outcome measures ranging from 1 to 5, a treatment effect of $-0.29$ can be regarded as a substantial negative effect, which is unlikely and in turn implies that our conclusion based on DR-ML is relatively robust to violations of the missing-at-random assumption. If we assume the outcome is missing at random, i.e., $\gamma_1 - \gamma_0 = 0$, the tipping point for selection bias is $\delta_1-\delta_0 = 0.26$ for DR-ML. Here, $\delta_1,\delta_0$ are the treatment effects among participants and nonparticipants, respectively. Taking the DR-ML point estimate $\widehat{\delta}_1 = 0.2$, this implies that the treatment effect among nonparticipants could be as low as $-0.06$ before altering the conclusion.
Compared to DR-ML, the tipping points for DR-PM are 0.35 and 0.21 for the above two scenarios. This suggests that DR-PM could be more sensitive to departure from the missing data assumptions than DR-ML, likely due to its larger variance. 

{
\bibliographystyle{apalike}
\bibliography{references}
}